# Smart Network Field Theory: The Technophysics of Blockchain and Deep Learning

Melanie Swan

Technology Theorist at Purdue University and Founder at Institute for Blockchain Studies
melanie@BlockchainStudies.org

Renato P. dos Santos

Blockchain Researcher at *ULBRA – The Lutheran University of Brazil*
renatopsantos@ulbra.edu.br

**Abstract:**

The aim of this paper is to propose a theoretical construct, *smart network field theory*, for the characterization, monitoring, and control of smart network systems. *Smart network systems* are intelligent autonomously-operating networks, a new form of global computational infrastructure that includes blockchains, deep learning, and autonomous-strike UAVs. These kinds of large-scale networks are a contemporary reality with thousands, millions, and billions of constituent elements, and entail a foundational and theoretically-robust model for their design and operation. Hence this work proposes *smart network field theory*, drawing from statistical physics, effective field theories, and model systems, for criticality detection and fleet-many item orchestration in smart network systems. Smart network field theory falls within the broader concern of *technophysics* (the application of physics to the study of technology), in which a key objective is deriving standardized methods for assessing system criticality and phase transition, and defining interim system structure between the levels of microscopic noise and macroscopic labels. The farther implications of this work include the possibility of recasting the P/NP computational complexity schema as one no longer based on traditional time (concurrency) and space constraints, due to the availability of smart network computational resources.

## Introduction

This paper proposes *smart network field theory* as a mechanism for the characterization, monitoring, and control of *smart networks* (intelligent autonomously-operating networks). The theory is developed qualitatively not algorithmically, with close association to two model systems, Cowan's Statistical Neural Field Theory and Wolynes's Spin Glass Model.

A *smart network field theory* is an effective field theory or any formal method for the characterization, monitoring, control of smart network systems such as blockchain economic networks and deep learning networks. There are different kinds of smart networks (Figure 4), and blockchain and deep learning are chosen as the focus for developing a field theory since they are among the most sophisticated, robust, and conceptually novel.

The term *field* is meant more analogically than literally (in the precise physical sense of an electromagnetic or gravitational field). In smart network field theory, field is referring to the ability to control fleet-many items as one unit (for example, as a field of neurons is controlled with a light switch in optogenetics). Field is also meant in the sense that a global operator (Hamiltonian) or variance principle (action or path integral) is derivable. The idea is that every point in a landscape has a value which may be calculated (per effective field theories in physics). The benefit of using operators and variational principles (to derive a "blockchain Hamiltonian" or a "perceptron path integral") is that a point value can be obtained which corresponds to a concurrency-based spatial and temporal configuration of an underlying dynamical system. Hence operators and variational principles, instantiated in smart network field theory, might be used to identify and manage critical points and phase transitions in smart network systems.

The current research extends ongoing work in *Technophysics* (the application of physics to the study of technology) (Figure 2). This includes the complexity theoretic analysis of blockchain consensus algorithms to assess system criticality such as chaosticity and flash crashes.[1,2] Other work theorizes smart networks,[3] articulating the smart network resource stack,[4] and aligning parallel formalisms in blockchain, deep learning, and black swan finance such as logistic regression, convexity, and programmable risk.[5] A special form of smart network field theory has been proposed for the management of medical nanorobots in the biological setting.[6] In deep learning, technophysics approaches have been used to facilitate system convergence with spin glass models, for energy landscape minimization (via Hamiltonian).[7] Deep learning has also been applied to the study of particle physics.[8] Complexity theory has been investigated for the control of another smart network system, drones or Unmanned Aerial Vehicles (UAVs), which exhibit undesirable emergent behavior such as thrashing, resource-starving, and phase change.[9]

There are four parts to this paper (Figure 1). Sections 1 and 2 introduce the concepts of technophysics, smart networks, and smart network field theory. Sections 3-5 develop the structure and requirements of smart network field theory, with statistical physics, effective field theories, and two model systems (Cowan's Statistical Neural Field Theory and Wolynes's Spin Glass Model). Sections 6-7 consider practical applications of the smart network field theory. Section 8 discusses the implications of smart network field theory for concurrency. The paper then considers risks and limitations, and concludes. A more detailed theoretical derivation of the model systems appears in the Appendices, and a Glossary is included.

**Figure 1.** Contents.

| |
|---|
| Section 1-2: Technophysics, Smart Networks, and Smart Network Field Theory |
| Section 3-5: Development of Smart Network Field Theory |
| Section 6-7: Practical Applications of Smart Network Field Theory |
| Section 8: Implications of Smart Network Field Theory for Concurrency |
| Appendix 1: Model Field Theory #1: Statistical Neural Field Theory (Cowan) |
| Appendix 2: Model Field Theory #2: Spin Glass Model (Wolynes) |

## Section 1: Technophysics and Smart Networks
### Technophysics
*Technophysics* is the application of physics concepts and methods to the domain of technology, particularly for the study of smart network technologies, algorithms, and computation.[6] The complex behavior and fast-paced emergence of smart network technologies such as blockchain economic networks and deep learning pattern recognition systems suggest that rigorous methods for their study is needed, and that physics would be an appropriate theoretical foundation. Figure 2 outlines the definition, scope, and research topics of Technophysics, and indicates the field's lineage by analogy to Biophysics and Econophysics. A key objective is deriving standardized methods for assessing complex system criticality and phase transition (for example, via a wider application of renormalization group theory), and defining interim system structure between the levels of microscopic noise and macroscopic aggregates.

**Figure 2.** Technophysics: Definition, Scope, and Research Agenda.

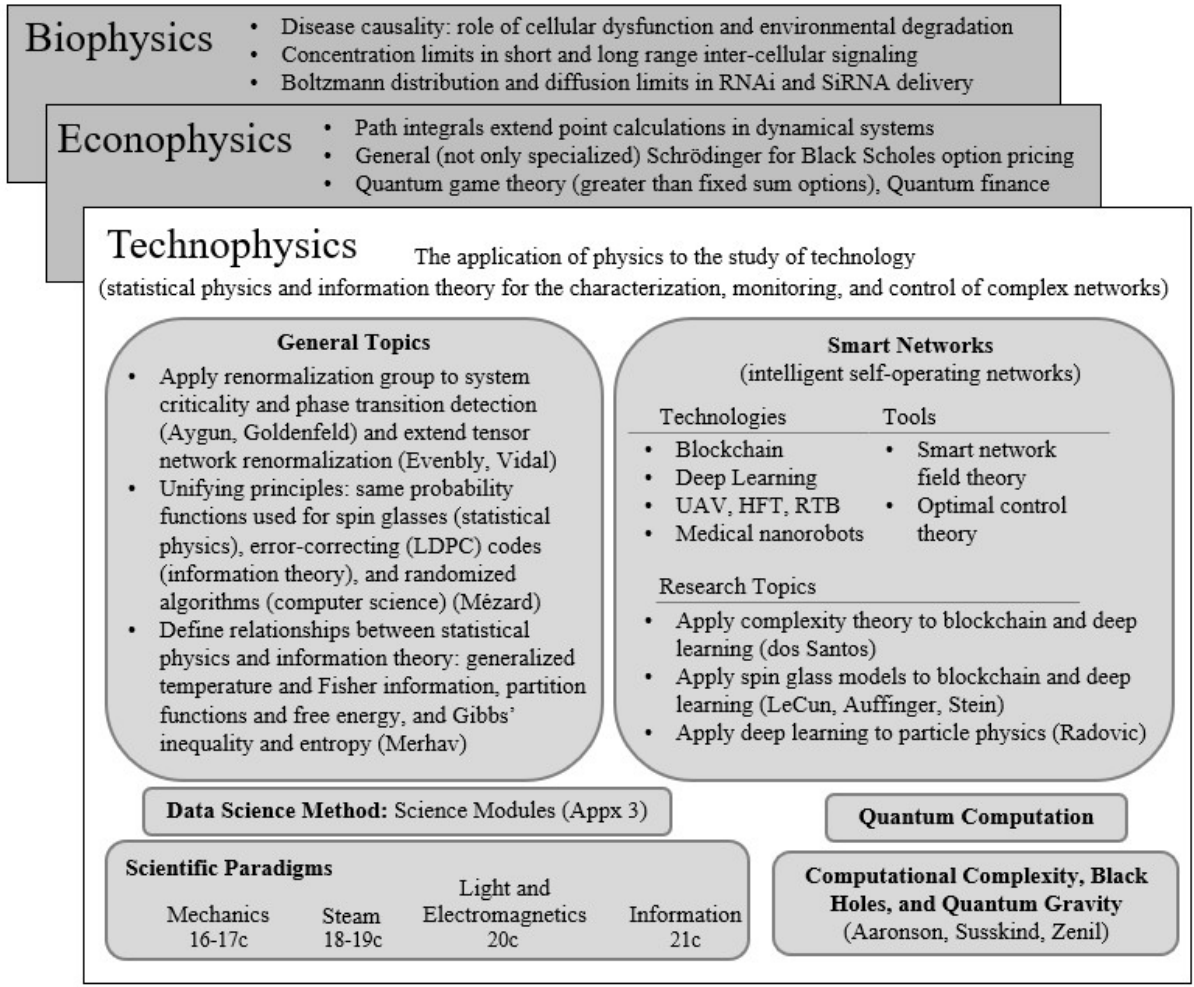

## Smart Networks

*Smart networks* are intelligent autonomous networks, a new form of global computational infrastructure in which intelligence is built directly into the software such that an increasing degree of autonomous operation is facilitated. More formally, smart networks are state machines that make probabilistic guesses about reality states of the world and act upon this basis. Whereas previously the hardware pipes and the network architecture were key focal points in network analysis, now it is the intelligent software that operates the networks. The notion of smart networks is configured in the conceptualization of there being two fundamentally two eras of network computing (Figure 3). Most progress to date, from mainframe to mobile, has concerned the transfer of basic information on simple networks. Now, however, in a second phase of network computing, a new paradigm is being inaugurated, smart networks, for the transfer of intelligent processing.

**Figure 3.** Two Eras of Network Computing: Simple Networks and Smart Networks.[3]

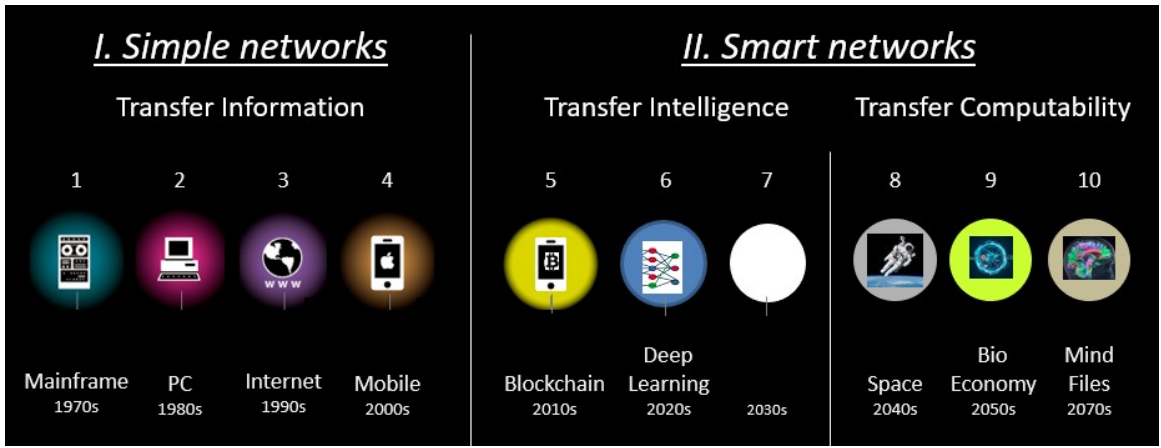

Blockchain distributed ledgers and deep learning systems are some of the most prominent examples of emerging smart network technologies (Figure 4). However, several other kinds of smart network technologies (autonomous software operating over networks) already exist and have been operating for some time. These include unmanned aerial vehicles (drones), particularly those that have autonomous strike capability,[10] programmatic or high-frequency trading (HFT) which now comprises 55 percent of U.S. equities trading volume,[11] the real-time bidding (RTB) market for advertising, and smart energy grids with automated load-rebalancing. Other emerging and envisioned smart network technologies include smart city Internet of Things (IoT) sensor ecologies, automated supply chain logistics networks, cloudmind industrial robotics coordination networks (cloud-connected smart machines),[12] personal robotic assistant networks, and space-spaced logistics networks.

Blockchains and deep learning networks are the focal smart network systems for the field theory development. Blockchains (transaction blocks cryptographically linked together) are one topology among others in the more general class of *distributed ledger technologies*. Blockchain-based verification is expensive, so alternative cryptographic structures have been proposed that might be more efficient. Another topology is "blockless blocks" in the form of directed acyclic graphs (DAGs). Emerging DAG projects include IOTA, Hashgraph, Byteball, and DAGCoin.[13] Efficiency is introduced in that IOTA is unmined, and consensus is established at the edge of the network, by a coordinator algorithm.[14] The coordinator algorithm uses a Curl hashing algorithm similar to the Hashcash proof of work algorithm used by Bitcoin,[15] for the same purpose of preventing spam and Sybil attacks. Nodes wanting to post new transactions to the network must first confirm a few other peer transactions before broadcasting their own. Smart network technologies may include different types of structural architecture, and also diverse operational objectives. For example, Ripple is notable as a species of blockchain network that provides an always-on live credit network (with over 29 million transactions since inception in 2012, half of which are related to maintaining open IOU credit links for future monetary transfer).[16]

**Figure 4.** Smart Networks Types and Operational Focus.

| | **Smart Network Species** | **Smart Network Operational Focus** | **Status** |
|---|---|---|---|
| 1 | Unmanned aerial vehicles (UAVs) | UAV drones with autonomous strike capability | E |
| 2 | Programmatic trading (high-frequency trading (HFT)) | Algorithmic trading (55% U.S. equities), auto-hedging | E |
| 3 | Real-time bidding (RTB) market for advertising | Automated digital advertising placement and inventory management | E |
| 4 | Energy smartgrids | Automatic electricity load-balancing & transfer on power grids | E |
| 5 | Blockchain economic networks | Transaction validation, confirmation, and recording | E/M |
| 6 | Deep learning networks | Object identification (IDtech[a]), pattern recognition, optimization | E/M |
| 7 | Smart City IoT sensor landscapes | Traffic navigation, data feeds, markets, and acquisition | M |
| 8 | Industrial robotics cloudminds | Industrial coordination (cloud-connected smart machines) | M |
| 9 | Supply chain logistics nets | Automated sourcing, ordering, shipping, receiving, payment | M |
| 10 | Personal robotic assistant nets | Personalization, backup, software updates, fleet coordination | F |
| 11 | Space: Aerial logistics rings | On-site resource provisioning and asynchronous communication | F |

*Legend: E: existing, M: emergent, F: future possibility*

### Deep Learning Chains

The diverse smart network technologies are emblematic of an emerging class of global computational infrastructure that increasingly acts autonomously based on intelligence built directly into the network operating software. Smart network technologies may be used in convergence. Some species of smart network technologies, such as blockchain and deep learning, are in a special class in that they are both smart network technologies themselves, and can serve as *control* technologies for other smart network systems. The functionality of blockchains and deep learning converges in the concept of deep learning chains. *Deep learning chains* are a smart network control technology with properties of both blockchain and deep learning: the secure automation, audit-log tracking, remunerability, and validated transaction execution of blockchains, and the object identification (IDtech[a]), pattern recognition, and optimization technology of deep learning.

Deep learning chains might be used to control other fleet-many internet-connected smart network technologies such as UAVs, autonomous driving fleets, medical nanorobots, and space-based asteroid mining rigs. For example, deep learning chains could be important in autonomous driving fleets for tracking what the vehicle does (blockchain), and for identifying objects in its driving field (deep learning). Deep learning chains could likewise apply to the body, as a smart network control technology for medical nanorobots, identifying pathogens (deep learning) and tracking and expunging them (blockchain). Likewise in supply chain automated receiving, deep learning chains could provide the integrated functionality of object recognition and validated transfer.[17] An example of deep learning chains in operation is Provenance's food supply chain traceability and attribute tracking system, which uses mobile apps, IoT sensors, blockchain, and machine learning. Sophisticated smart network control technologies such as deep learning chains require a strong theoretical model for their design and operation, and this is one motivation for the smart network field theory development proposed by this work.

---

[a] IDtech: identification technology: the technological capability to recognize objects; similar to FinTech, RegTech, TradeTech, and HealthTech; technologies that digitize, standardize, and automate their respective domains

## Section 2: Smart Network Field Theory

Large-scale networks are a feature of contemporary reality. They are complex systems comprised of thousands, millions, and billions of elements, and thus require a smart network field theory or other similar mechanism for the automated characterization, monitoring, and control of their activity. A theoretically-grounded model is necessary, and a smart network field theory based on statistical physics, effective field theories, and model systems is proposed here. Other models might likewise be developed for the analysis and control of large-network systems. A s*mart network field theory* is a field theory for the characterization, monitoring, and control of smart networks systems, particularly for criticality detection and fleet-many item orchestration.

### *Requirements: Characterize, Monitor, and Control*

The purpose of having one or more smart network field theories is for the characterization, monitoring, and control of smart network systems. The first objective is characterization. It is necessary to develop standard indicators and metrics to easily identify specific behaviors in smart network systems as they evolve and possibly grow in scalability and deployment. Both positive (emergent innovations) and negative (flash crash) behaviors should be assessed.

The second objective of a smart network field theory is to provide monitoring, at both the individual element and overall system level, of current and evolving behavior. Monitoring pertains to smart network operations currently unfolding, and also farther future needs. In the farther future, *deep thinkers* (advanced deep learning systems) might go online. Although deep learning networks are currently isolated and restricted to certain computational infrastructures, it is not unimaginable that learning algorithms might be introduced to the internet. A *Deep Thinkers Registry* could be an obvious safeguard for tracking an entity's activity, with possible annual review by a Computational Ethics Review Board for continued licensing. This is a far future example, but demonstrates the intended extensibility of smart network field theories, and the uncertain future situations that they might help orchestrate.

The third objective of a smart network field theory is control, both for the coordination of fleet-many items, and for predictive risk management. The coordination of fleet-many items is an obvious automation economy benefit of smart network technologies. This is the ability to coordinate fleet-many items in any kind of internet-connected smart network system, which could include autonomous vehicles, drones, blockchain p2p nodes, deep learning perceptrons, smart city IoT sensor landscapes, home-based social robots, medical nanorobots, and supply chain shipment-receiving. The longer-term range of deployment of smart network technologies could extend to the very small, the cellular domains of the body, and the very large, such as terraforming in space.

### *Applications: Fleet-many Coordination and System Criticality Detection*

One of the most practical applications of a smart network field theory is the automated coordination of fleet-many items. Another is predictive risk management in being able to detect and possibly avert system criticality such as potential phase transitions. A crucial use of a smart network field theory is for the predictive risk management of system criticality. It is important to have a mathematical and theoretical basis for understanding smart networks so that the predict critical points and phase transitions may be predictively managed to the extent possible. What could constitute criticality and phase transitions in smart networks is both expected and emergent situations such as financial contagion, network security, and electromagnetic pulses (either malicious or unintentional).

Smart network field theories could also be useful in the well-formed design of smart network systems. A smart network field theory provides a formal scientific basis for studying smart networks as new technological objects in the contemporary world, particularly since smart networks are a nascent, evolving, and high-impact situation. A general form of the smart network field theory is developed here. Special forms could be elaborated for specific applications, for example, for control systems in contexts of the very-small, brain-computer interface (BCI) implants,[18] and the very-large, space-based mining and secure asynchronous communication.[19]

## Section 3: Smart Network Field Theory Development

### Methods

This section develops the smart network field theory. A *smart network field theory* is any formal method for the characterization, monitoring, control of *smart network* systems such as blockchains and deep learning networks. Although there are different kinds of smart networks (Figure 4), blockchain and deep learning are the focus for developing a field theory because they are the most sophisticated, robust, and conceptually novel.

The term *field* is meant more analogically than literally (in the physical sense), and likewise other terms subsequently invoked in this work such as *temperature* and *pressure*, which also have precise meanings in the physical context. These terms are applied conceptually as to the purpose and function they are serving in physical systems, which is then extended to the context of smart network field theory development.

There are two primary meanings of *field* in the more conceptual smart network field theory sense. First and most generally, field is referring to the ability to control multiple items as one unit. The requisite functionality is to manage fleet-many items. One idea is to control them as a field, where "field" might be dynamically-defined based on energy, probability, gradients, or other parameters. The concept of field might be used to coordinate thousands and millions of constituent elements (such as blockchain p2p nodes or deep learning perceptrons). An example of an existing smart network field operation is *optogenetics* (in which neurons express a protein that makes their electrical activity controllable by light).[20] Optogenetics is a "light switch for a field of neurons" in that it conveys the ability to turn on or off a field of neurons at once. Thus, a smart network field theory is constituted in that optogenetically-enabled cells are controlled as a field as opposed to individually.[6]

The second meaning of *field* in smart network field theories is as follows. A field might refer to the situation in which each element in a system has its own measure and contribution to overall network activity (and can be used to calculate a Hamiltonian or other system composite measure). This concept of field (from effective field theory development in physics) suggests that every point in a landscape has a value which may be calculated.

Smart network field theories may be structured in scalar, vector, and tensor terms (meant in the usual way). A tensor is a complex mathematical object (a geometric object that maps functions and interacts with other mathematical objects). The simplest tensors are scalars (zero rank tensors) and vectors (first rank tensors). A scalar is a tensor with magnitude but no direction (a zero-rank point), and is described by one number. Mass and temperature are scalars. A vector is a tensor with magnitude and direction (a first-rank line). A vector is often represented as an arrow, and defined with respect to a coordinate system. A tensor is a multi-dimensional structure, such as a matrix. Tensors are seen in the context of deep learning. Google's TPUs (tensor processing units) and TensorFlow software use *tensor* in the sense conducting very-fast matrix multiplications. A tensor is a multi-dimensional matrix, and TPUs and TensorFlow are fast because they flow through matrix multiplications directly without storing intermediate values in memory.

**Theoretical Approach to Smart Network Field Theory Development**

Statistical mechanics and effective field theories are employed as the theoretical domains to underpin *smart network field theory* development. Two model systems further demonstrate these principles (discussed in Section 4). A key requirement for a smart network field theory is that it can be used to manage across diverse system scales. A smart network field theory should be able to "identify macroscopic smoothness from microscopic noise" as complexity theory articulates.[21] Statistical mechanics and effective field are therefore selected as two methods for linking multiple dimensions within systems.

*Statistical Mechanics, Probability, and Thermodynamics*

> "*All of physics in my view, will be seen someday to follow the pattern of thermodynamics and statistical mechanics*" – John Archibald Wheeler, 1983, 398[22]

Statistical physics is an important generalized method, rooted in probability, for linking microscopic noise to macroscopic labels.[23] Smart networks too are fundamentally based in probability. Smart network technologies such as blockchain and deep learning are probabilistic state machines that coordinate thousands to millions of constituent elements (processing nodes, which can be seen as being analogous to particles) to make high-probability guesses about reality states of the world. Thus, statistical mechanical field theories could be good models for the formalization of smart networks.

Maxwell was among the first to suggest the application of probability as a general method for the study of the science of the very small. Statistical mechanics was likewise intended as a general method based on probability in its early development by Gibbs, following on from classical mechanics.[24] Statistical mechanics was to address all aspects of mechanical systems, at both the microscopic and macroscopic levels, for example, explaining transitions between gaseous and non-gaseous states.

The thermodynamics aspect of statistical mechanics is also an important and obvious formulation for the study of smart networks. Concepts such as work, heat, and energy have analogs between thermodynamical systems and smart networks. Blockchains have work in the form of proof of work consensus algorithms being a primary mechanism for providing network security and updating the ledger balances of the distributed computing system. Deep learning networks also perform work in the sense of conducting work to run an operating cycle to derive a predictive classification system of data. The network expounds significant system resources to feed forward and backpropagate in order to optimize trial-and-error guesses about the weighting of relevant abstracted feature sets to classify new data.

Further, in blockchains, cryptographic security is based on thermodynamics and entropy (it is unknown which machine at random will confirm the next transaction block). The network is provably unchangeable through thermodynamic immutability, in the sense of being able to compute the amount of energy (in joules) that would be required to change the ledger.[25]

The laws of thermodynamics hold in the sense that it would not be possible to transfer free energy without a corresponding energy expenditure in the form of entropy that the system must spend.

Macroscopic statistical quantities thus make sense in that there is a state of thermodynamical equilibrium in both blockchain systems and deep learning networks. In blockchain systems, equilibrium may be constituted different ways, including by a match between the replicability of information and the competition among peers writing to the ledger.[26] In deep learning networks, the system is likewise at equilibrium, finding an optimized solution after some number of forward and backward propagations.[27]

### *Applicability: Physical System Status*

Since Statistical Mechanics and Effective Field Theories are developed in the context of particle and matter-based physical systems, a question arises about the physical status of smart network systems and the extent to which physical theories would apply. Blockchains are physical systems, about 10,000 nodes on average hosting the transaction ledger for Bitcoin (9,892[b]) and Ethereum (14,098[c]). Deep learning networks also have a physical basis in that they run on large dedicated hardware systems (NVIDIA GPU networks and Google TPU clusters (tensor processing units)). However, deep learning networks are more virtualized. The operation of the network is as computation graph representations of a network. The conclusion at present would be that the field theory methods might apply more readily to blockchain systems. They might likewise apply to deep learning systems, but with more contextualization, especially if deep learning techniques were to go online in a more distributed format. For both smart network systems, a technophysics research opportunity exists to apply physics models such as statistical mechanics and effective field theory development to define formalisms more precisely.

### *Effective Field Theories*

The other arm of smart network field theory development is effective field theories, a known technique used in basic physics, particularly quantum mechanics and materials science.[28] An *effective field theory* is a type of approximation, or effective theory, that explains an underlying domain in simpler terms. Effective field theories are useful because they span classical and quantum domains, and more generally, different levels in systems with phase transitions. One of the most prominent examples of an effective field theory is Quantum Field Theory (QFT). QFT extends Quantum Mechanics to descend down a system level and incorporate the energetic fields in which particles behave. Particles are reinterpreted as states of a quantized field.[29]

It is easier to measure different aspects of systems at different scales. For example, computing the energy spectrum of the Hamiltonian in Quantum Mechanics is straightforward, but much more difficult at the QFT level because there is too much data about particle movement. Likewise, the elliptical orbits of the planets are more easily calculated with Newtonian gravity than with general relativity. An effective field theory gives the ability to focus on a particular scale of a system, emphasizing one aspect while limiting others.[30] The objective is to find the simplest framework that captures the essential physics of the target area. For example, if there were interest in the lighter particles (such as bottom quarks), the heavier particles (e.g. z-bosons and w-bosons) could be eliminated from the model.

---

[b] https://bitnodes.21.co (accessed September 18, 2018)
[c] https://www.ethernodes.org/network/1 (accessed September 18, 2018)

In multi-level systems, identifying a macroscopic term corresponding to microscopic behavior is a key challenge. The analogs to the temperature and pressure terms arising from a room of moving particles in a model system are not always clear. An effective field theory is a formal process that can be used to identity a system's temperature term and other system-level parameters. Effective field theories are similar to the renormalization concept (mathematically shifting to a different level of system scale to focus on a parameter of interest). Deriving an **effective field theory** involves the steps outlined in Figure 5: defining the system, the elements, interactions, and levels or scales are of interest (degrees of freedom), what can be emphasized and eliminated in analyzing the system, and the relevant quantity that can be averaged over the system to produce a temperature-type term.

**Figure 5**. Steps to articulate an Effective Field Theory.

| |
|---|
| 1. Characterize the system: the overall scope, shape, and levels of the system |
| 2. Identify the constituent elements of the system, the kinds of interactions between them, and the overall quantities in the system |
| 3. Identity the *degrees of freedom* (the aspects of the system that matter for the problem of study), and the irrelevant substructure that can be ignored |
| 4. Articulate the mathematics to measure the system, for example averaging the underlying behavior in to derive a simplified model with a global term such as a Hamiltonian or Lagrangian |

The crucial aspect is to identify what is important to measure in the system, for example the propagation, symmetry, range of system states, or available potential energy. Applying the effective field theory technique to the smart network context, the idea is to consider the different levels and dimensions of the system, and identify the elements, interactions, and relevant quantities to calculate in order to obtain the system behavior. For example, in a biological neural network, the system-wide quantity of interest might be the spiking activation (the threshold at which neurons fire), and other data would be superfluous.

**Section 4: Two Model Field Theory Systems**

This section considers two existing field theory systems for the development of the smart network field theory. The two model systems are Cowan's Statistical Neural Field Theory for describing brain activity,[31] and Wolynes's Spin Glass Model,[32] extended with LeCun's theoretical work in deep learning systems,[33] to specify how protein-folding and deep learning systems converge on a solution. A summary of the main themes is presented here. A more detailed version appears in the Appendices with diagrams highlighting the specific steps in the theory derivation, including the specific problems tackled, the physics-based techniques applied, and the results obtained. Both theories start with a basic characterization of the system, using the Wilson-Cowan equations as a mean field theory (Cowan), and the Random Energy Model (Wolynes). Then the main analytical models are proposed: Cowan's Statistical Neural Field Theory and Wolynes's Spin Glass Model (Figure 6). The theories are analytically solvable, and thus might be applied to the formal development of smart network field theories.

**Figure 6**. Field Theory Model Systems: Characterization and Analytical Model.

| Model System | System Characterization | System Analytical Model |
|---|---|---|
| 1. Cowan | Wilson Cowan equations (mean field theory) | Cowan Statistical Neural Field Theory |
| 2. Wolynes | Random Energy Model | Spin Glass Model |

**Model Field Theory #1: Statistical Neural Field Theory (Cowan)**

First, to describe the general equilibrium state of the system (a biological neural network; i.e. a brain), a mean field theory using the Wilson-Cowan equations is articulated. Second, to model system criticality and phase transition, a statistical field theory is proposed (Cowan's Statistical Neural Field Theory). The statistical field theory incorporates the effects of correlations and fluctuations in the dynamical non-equilibrium system. The overall structure of the field theory is to define the dynamical system, its state transitions, how action occurs within the system and evolves, and system criticality and phase transition, and how they might be controlled.

The system is formulated as a Markov process (random walks, random fields) to derive path integrals for overall system measures (correlation (distance) functions, moment-generating (probability distribution) function, and action). The action (an abstract quantity describing the aggregate motion of the system such as neural firing threshold) is derived as the path integral of the Lagrangian (the overall configuration of network fields, particles, and states).

With the path integral, the next goal is writing a master evolution equation for the system. A single random walk is generalized to random walks in a continuum, to frame the problem as a Markov random field. A neural dynamics of the system is obtained by quantizing the neuron into three states, quiescent, active, and refractory (QAR). A two-state model (quiescent and active (QA)) could also be used. Simplifications are applied such as Wick's theorem, Gell-Mann's 3x3 matrices, and eigenfunction ladder operators. An algebraic version of the master evolution equation is specified which describes the QAR state transition diagrams of the neural network.

Having described the dynamical system, to articulate how action occurs within the system, a neural state vector (a probability state vector) is derived. This is possible because the algebraic equation is a number density operator that can be used to count the number of neurons in the different QAR states, and the total excitation coming onto the network (with a Hamiltonian) as a network quantity. Euclidean field theory (a modern form of statistical mechanics) is applied to derive a Wiener-Feynman path integral (one that is mathematically well-defined by using imaginary time (unlike the original Feynman path integral which is not easily computable)). This allows a simplification to write the system with only the spike Hamiltonian instead of the full Hamiltonian (the Hamiltonian indicating the spiking threshold of the system).

Having described the dynamical system and how action occurs within it, next is defining a mathematical model of system evolution, for which coherent states are used. Coherent states are functions that do not change their state under the action of the evolution of the Schrödinger wave equation. A theory of spiking action on a neural network is obtained using coherent states, applying the U(1) symmetry (rather than that of the more complicated Lie algebra topology group SU(3)). A theory of spiking action explains system evolution.

However, this is a linearized spiking model, so to specify a non-linear spiking model, perturbation techniques (the renormalization group) are used to define a renormalized action. The issue is that there are multiple time and space scales in non-linear systems, so a renormalized action that takes into account new critical points at longer time and length scales is necessary.

The renormalized action is the same as the action in Reggeon Field Theory (which characterizes strong interactions in high-energy particle physics). The behavior of a Reggeon Field Theory system corresponds to both branching and propagation (the crucial criticality in a neural system). Such systems have a universal non-equilibrium phase transition, in a class called directed percolation. System criticality is thus articulated in terms of directed percolated (unidirectional) phase transitions. To control the system, the action of an optimal control theory can be expressed on the dynamical system as a path integral. A summary is in Figure 7.

**Figure 7.** The Statistical Neural Field Theory (Cowan).

| Key findings from Cowan's Statistical Neural Field Theory |
| --- |
| • A statistical field theory is needed to model the effects of correlations and fluctuations in dynamical non-equilibrium systems |
| • A field theory is specified to identify system state transitions, evolution, and trigger points (a *field theory* is defined as a set of functions to assess system state and criticality) |
| • The problem is formulated such that path integrals can be derived for overall system measures (functions and point values corresponding to underlying system configurations) |
| • The problem (system) is constantly simplified with the application of various physics methods |

## Model Field Theory #2: The Spin Glass Model (Wolynes)

First, the problem at hand is formulated as an energy function, for example towards the practical solution of problems in protein-folding and deep learning networks. A Random Energy Model is used to obtain a general characterization of the energy landscape. Second, a Spin Glass Model (a more complicated extension of the Random Energy Model) is applied to derive an exact solution for the system.

A Random Energy Model is used to make random (Gaussian) guesses about the probability distribution of the overall energy available in the system. However, since these are only random guesses, the system might not ever produce order, meaning converge on a solution. In protein folding, this is called the Levinthal Paradox. The paradox is that proteins fold in a matter of nanoseconds, so they cannot possibly be folding randomly because it would take longer than the known history of the universe for all possible permutations to be tried.[34]

Therefore, in order to direct the energy landscape into a solution (such as obtaining a folded protein or a facial recognition), the Spin Glass Model is introduced. A *spin glass* is a disordered magnet that is a metastable system with half of its molecular bonds oriented in the direction of spin-up, and the other half in the direction of spin-down. The term *glass* is meant as an analogy between the magnetic disorder in the spin glass and the positional disorder in a conventional glass. Conventional glasses have an irregular atomic structure (unlike crystals which have a uniform atomic structure), and even though glasses are solid enough for everyday use, are technically super-cooled liquids. The benefit of using a spin glass model is that the spins can be relaxed into real numbers, so that the system can be solved analytically. This means that the random energy function can be instantiated as a Hamiltonian (a stratified function that is a weighted probabilistic sum of the total energy in the system) that can be solved.

The goal of the spin glass function formulation is to have the energy landscape converge on a solution, here, producing a folded protein, or an error-minimized accurately-classified set of test data in a deep learning network. Wolynes's Spin Glass Model is summarized in Figure 8. Spin glass models are applied to neural networks in the brain,[35] and likewise deep learning neural networks. Spin glass models are applied to deep learning systems for optimization, in order to create efficient *loss functions* that reduce combinatorial complexity and minimize cross entropy. Loss optimization is formulated as an energy function with Hamiltonians. System energy is calculated in the ground state and subsequent tiers of progression of the system,[36] which is interpreted as a transition from a glass to a spin glass system (a glass transition), i.e. a solution to a smart network loss optimization equation.[7]

**Figure 8.** The Spin Glass Model (Wolynes).

| Key findings from Wolynes's Spin Glass Model |
| --- |
| • The problem is formulated as an energy function<br>• First, a Random Energy Model makes random guesses about the total energy in the system<br>• Second, a Spin Glass Model is used to get the energy landscape to converge on a solution. Flat glassy landscapes are overcome by producing a glass transition, a rugged, convex funnel in the energy landscape that produced a point solution<br>• The system is analyzed with *energetics* (the information-theoretic trade-off between energy and entropy). The phase transition occurs at a "critical temperature," at the moment when the system entropy and energy converge (Appendix 2, Figure A2a)<br>• System performance is optimized by minimizing loss and cross-entropy (efficient information encoding) modeled as an energy function convergence or glass transition |

**Application of the Model Field Theory Systems to Smart Networks**

The theoretical foundations of smart network field theory development in statistical mechanics and effective field theories are joined with the model systems and applied to smart networks to instantiate smart network field theories more concretely (Figure 9). The crucial elements of the field theories are organized into columns. These include whatever in the system is constituting the particles, interactions, temperature term, Hamiltonian term, system-spanning mechanism, threshold trigger, what constitutes system criticality and phase transition, and the optimal control mechanism.

The rows articulate these aspects in the different systems studied. The side-by-side comparison indicates that while smart network systems do not have exact equivalents to traditional field theories, the structure is very similar and field theories could be a good model for studying, characterizing, monitoring, and controlling smart network systems with physics-based principles. The *temperature term* is understood to be providing a descriptive summary term at the overall systems-level. The *Hamiltonian term* is a point value that corresponds to an entire configuration of elements in the underlying system, a formalized measure of capacity which identifies the number or percent of system nodes that are available to provide certain services.

**Figure 9**. Effective Field Theory Diagram: Model Systems and Smart Network Systems.

| Network Species | Particles | Interactions | Field Theory Elements | | | Threshold Trigger | Phase Transition | Optimal Control Mechanism |
|---|---|---|---|---|---|---|---|---|
| | | | 1. Temperature Term | 2. Hamiltonian Term | 3. Scale-spanning | | | |
| *Model Systems* | | | | | | | | |
| Traditional physical system | Pollen molecules in water | Particle vibration, Brownian motion | Temperature, pressure, volume, amount | Lagrangian, action | Measurement | Temperature level | Gas: ice, liquid, vapor | Temperature lever |
| Neural network in the brain | Neurons | Signaling, firing cascades | Signaling potentiation | Firing rate function | Moment-generating function | Spiking function | Signaling cascade, DP Phase Transition | Path integral of the OCT as an Action |
| Protein Folding | Nucleotides | Structural contacts | Polymer folding capacity | Hamiltonian | Conformation signaling | Rugged energy funnel | Folded protein | Polymer dynamics |
| Deep Learning #1 (model system) | Artificial neurons (perceptrons) | Loss minimization | Convergence, errors | Hamiltonian | Glass transition | System energy | Convergence | System architecture |
| *Smart Network Systems* | | | | | | | | |
| Blockchain | Peer-to-peer nodes | Confirmation, transfer, exchange | Hash rate | Blockchain Hamiltonian (quantity function) | Merkle Tree (hashes) | Algorithmic trust, information liquidity | 51% or DDoS attack, system fork | Curve smoothness |
| Deep Learning #2 (separate analysis) | Perceptrons | Feedforward back-propagation node weight-bias adjustment | Error rate | Activation (ReLU), Regularization | Long short term memory (LSTM) | Object IDtech, pattern recognition, optimization | Endless loop, no stopping | Adversarial net, dark knowledge methods |
| UAV | Vehicles/drones | Positional communication | Degree of thrashing | Resource-starved system | Information sharing | Deadlock | (Anti) coordination | Collective intelligence |
| Programmatic trading (HFT) | Trades, trading agents | Trading, smart contract posting | Volatility | Liquidity | Sentiment broadcast | Circuit breakers | Flash crash | Vega hedging |
| Medical Nanorobots | Nanorobots | On-demand grid formation | Healthy cells, homeostasis | Hamiltonian | Cell-organ-tissue signals | Disease | Pathology resolution | Cure, enhancement |

*Temperature term: a consolidated measure comprising a microscopic level of activity at one or more tiers down in a system*
*Hamiltonian term: an operator or variational principle producing a point value of an underlying system configuration*
*Scale-spanning: mechanism for spanning multiple system levels, including optimizing intervention where the function curve is the smoothest*

**Section 5: Elements of Smart Network Field Theory**

This section proposes three basic elements as a minimal configuration for smart network field theories (Figure 10), based on the theoretical foundations of statistical physics, effective field theories, and model systems.

**Figure 10**. Smart Network Field Theory: Minimal Elements.

| | Element | Function |
|---|---|---|
| 1 | Temperature term | Macroscopic label for microscopic behavior |
| 2 | Operators (Hamiltonian, Lagrangian) and Variational Principles (Action) | Point value (real number) for a dynamical system configuration |
| 3 | Portability across System Scales | Indication of optimal level for system engagement |

**Temperature Term**

It was previously distinguished that physics terms may have two meanings in the context of smart network field theory development, one that is precise and analytical, and one that is conceptual and analogical. These terms, such as *temperature* and *pressure*, might be used different ways in smart network systems: literally as a physical quantity, conceptually as a system control lever, and analytically as a system assessment parameter.

Before discussing these uses, the first point to clarify is the justification for a temperature term in smart network systems. The claim is that any kind of system that can be described with statistical mechanics, including smart networks, may have different kinds of probability distributions, including Maxwell-Boltzmann distributions (probability distributions of a quantity) and ensemble distributions (probability distributions over all possible states of a system). The probability distributions allow a macroscopic term to be formalized. This means that any random conserved energy phenomenon at the microscopic level of a system will have a macroscopic term capturing the aggregate activity (e.g. temperature, pressure, etc.). A *temperature term* occurs naturally in any system in which there is a conserved quantity, which could include energy or system state, and therefore apply to smart networks.

Temperature is first used literally, in the operational sense as a physical quantity that expresses hot and cold. Temperature is a proportional measure of the average kinetic energy of the random motions of the constituent particles of matter (such as atoms and molecules) in a system. Smart network systems are physical, and thus temperature as heat makes sense to infrastructure operators and anyone else engaged in the physicality of the system such as for design purposes. In smart networks, one form of the *temperature term* is a quantitative measure of the overall system, such as blockchain hash rate and deep learning system error rate.

Temperature is also used conceptually, as the temperature term is based on theoretical arguments from thermodynamics, kinetic theory, and quantum mechanics A temperature or pressure term links the levels in a system between a microscopic and a macroscopic domain. Temperature is the consolidated measure that describes the movement of all of the particles in a room. Likewise, in a field theory, a temperature term is a consolidated measure that comprises all of the activity of a microscopic level at one or more tiers down in the system. The temperature term might be employed as a control lever for the system.

The third use of a temperature term in a statistical mechanical system is as a system assessment parameter. In deep learning systems, examples of a *temperature term* are regularization and weight constraints in the system activation function. The sum of the weights on the node activations comprises a temperature term that can be used to assess or control the system. Overall, the temperature term can be thought of as a valve for controlling a system-level quantity or resource, setting the scale of the "energy" or capacity in the system, and might be increased or decreased as a control and regulation mechanism. An emerging layer of network resources (such as algorithmic trust, information attributes, liquidity, and error correction) might be analogs to *energy* in a traditional physical system and managed similarly as levers for system-level regulation.

As a practical example, temperature terms are macroscopic measures that are chosen for being informative. For example, if the pressure rises above some working range or the temperature drops below it, it is an indication of some possibly dangerous malfunction of the system that must be overseen. An informative measure for blockchain systems would be one that warned, for example, that a 51% or DDoS attack was forming, or that there are so few online nodes that the security of the network is compromised.

**Operators (Hamiltonian, Lagrangian) and Variational Principles (Action, Path Integral)**
The use of a temperature term highlights the theme of having higher-level levers in a system that correspond to a whole class of underlying activity. Whereas temperature and pressure are macroscopic labels for one set of microscopic phenomena, another set of levers is available to capture the dynamics of the system, in the form of operators and variational principles. Operators and variational principles are terms that track the variability of certain quantities in a system. In the traditional physics setting, there are energy, momentum, and position operators, among others. Energy operators (which measure the total amount of energy in the system over time) may be the most familiar, the Hamiltonian in quantum mechanical systems and the Lagrangian in classical systems. The action is a path integral of the operator (a summation over all paths or trajectories of the system), from which the equations of motion of the system can be derived. For example, in Cowan's statistical neural field theory, the action is a path integral of the Lagrangian (the overall configuration of possible states), which is used to construct the moment-generating function, the threshold at which neural spiking (signaling) occurs in the brain.

The benefit of using operators and variational principles is that a real number can be obtained, a point value that corresponds to a configuration of the underlying system. These functions assign a real number to a configuration of particles or fields in a dynamical system, including temporal and spatial parameters. Therefore, operators and variational principles might be used to identify and manage critical points and phase transitions in smart network systems. For example, one way that a *smart network Hamiltonian* might be employed is to measure the point value of system capacity. This could be the number or percent of system nodes that are available to provide a certain service. The smart network Hamiltonian could indicate quantitative measures of system capacity.

**Portability across System Scales and Optimal System Engagement**
A key objective of a field theory is to be able to move portably across system scales. Smart network technologies (such as blockchain and deep learning) are not just one microscopic and macroscopic level, but rather may extend many levels up and down, both quantitatively, and qualitatively. Two examples of multi-level quantitative spanning in blockchains are first that U.S. Treasury officials track dollar bills at twenty different levels of detail. In a blockchain system, arbitrarily-many levels could be rolled up with a Merkle tree structure, as the entire Bitcoin blockchain (currently over 500,000 transaction blocks) can be called with one composite value, a single hash code or Merkle root. Controlled substance pharmaceutical inventories could be similarly tracked in blockchain-based digital asset registries and aggregated at any level of detail, on a hospital, state, national, or international basis. It might be expected that eventually all assets might be tracked with digital inventories and blockchain systems.

Turning to deep learning networks, there are examples of spanning both quantitative and qualitative system levels. Deep learning networks have five to seven layers of millions of processing nodes on average now, and different ways of abstracting these into higher-level feature sets. Convolutional networks combine simple features (a jaw line) into more complex feature sets (a face) for image recognition, and recurrent networks identify relevant sequences from data in speech recognition, using LSTM (long short term memory to flag what to remember and what to forget). Adversarial Nets (two networks pitted against each other)[37] and Dark Knowledge (compression systems)[38] are examples of higher-level roll-ups that are qualitative (i.e. different in kind not only magnitude). Applying machine learning to the better design of deep learning systems themselves is an example of recursive and qualitative system level spanning, using machine learned instead of hand-coded optimization algorithms,[39] and to find new topologies of deep learning architectures.[40]

Recognizing that smart networks are multi-level systems, the question arises as to where to optimally engage the system for certain operations. Smart network field theory can be used to calculate where the curve of a given function is the smoothest.[41] The field theory can determine at which derivative or integral level in the system the curve of a certain function is the smoothest. For example, in another model system, financial market options trading, there are four system levels: price, delta, gamma, and vega. The terms are partial derivatives of price; delta is the rate of change of price, gamma is the rate of change of delta, and vega is the rate of change of gamma. In this system, the strongest signal is at vega (the curve is the smoothest), and many traders engage the market at this level.[42] In smart network systems, partial derivatives of the Hamiltonian term could be taken as one means of determining where the curve is the smoothest.

## Section 6: Practical Applications of Smart Network Field Theory

This section considers the practical applications of smart network field theories, including the network nodes, states, actions, and phase transitions of the system.

### Nodes (Particles)

In smart networks, the *particles* are the constituent network elements. In blockchains, this is the p2p (peer to peer) nodes, and in deep learning networks, the perceptron processing units. In blockchain networks, the core unit is the p2p node. This is the flat-hierarchy distributed p2p nodes that offer fee-based services for activities such as transaction validation and recording (mining). There are also other units in blockchain networks such as wallets, transactions, digital assets, and digital asset inventories.

In deep learning networks, the constituent units are perceptron nodes. Perceptrons are modular units, organized in a graph-theoretic structure. The computation graph has nodes and edges. Input values enter the processing node from the input edges, are processed by logic in the node, and exit as an output value on the output edge, which is used as the input value for a downstream node. Thousands and millions of nodes are organized Lego-like into cascading layers of processing (five to seven layers of thousands or millions of nodes on average).

### Node States

The nodes, units, or particles in the system may have different states and values. For example, wallets in a blockchain network may have a ledger balance (e.g. an amount of money, contract status, or program memory state). Wallet states are analogous to the neuronal states in the Cowan theory. Smart networks are state transition machines (systems that track individual and global state changes). The state values might be binary (quantal; discrete), or continuous, depending on the modeling technique. A field-theoretic model can expand in complexity. Having continuous state values would require a more complicated configuration, and eventually introduce operators instead of merely point functions.

Nodes have states. In the Cowan model system, biological neurons might be in one of two or three different states (quiescent or active (QA), or also refractory (QAR)). In blockchain systems, the same two-state structure could be used initially, to model nodes that are either broadcasting or quiescent (BQ).[1,2] In deep learning networks, different node states might likewise be identified. These could include a basic measure of whether the node is processing or finished (PF), or more complicated measures such as a maximum or minimum error for this node, or a continuous value for error-maxing contribution. The system may already be set up to operate by calculating the maximum error contributed by each node. In deep learning systems, the softmax function is already a field-theoretic application of a renormalization function. Similar to the spin glass model's relaxation of spin values to real numbers for calculation, the *softmax* function exponentiates a set of numbers, and rescales them so that they can sum to one, allowing a state interpretation based on probability. The general idea is that field-theoretic formulations such as the softmax function allow deep learning networks to be evaluated with the Hamiltonian concept of a system-wide probabilistic measure.

**Node Action (measured with a Temperature Term or Hamiltonian-type Term)**
The softmax function is one example of a Hamiltonian-type operator in a deep learning smart network which is used to measure an overall system quantity, in the case of softmax, node error contributions. More generally, the idea is to identify system-level quantities that would be the target of a smart network field theory, and to create a Hamiltonian-type term to measure them. One of the most basic system quantities to measure could be system capacity, namely the overall capacity of nodes to provide services. For example, in a blockchain network, a Hamiltonian-type operator could be used to sum the probability of nodes being in the broadcast state of providing or being able to provide certain services.

A temperature term could be defined as the quantitative amount of network resource availability (expressed as a probability percentage, or as abstracted into a fixed metric like temperature that has meaning within the system). The probability of network resource availability could be calculated per the weighted probability of certain nodes being in certain states (e.g. Mining On (Y-N), Ledger-Hosting (Y-N), Credit Transfer (Y-N)). In deep learning, the same concept of aggregate node states measured as a global system variable could be used, for example, to measure system quantities such as error contribution, feature abstraction levels, and degree of sequentiality (in an LSTM RNN (long short term memory recurrent neural net). Smart network field theories are intended as a general purpose tool that could be used to measure a variety of aggregate network resources. For example, there could be the concept of "Cowan for Ripple," meaning the Cowan sum of the network node probability of credit availability.

Smart network resources (quantities) might be modeled as binary availabilities, like the p2p network node status BQ (broadcast or quiescent, just like the binary neuron system QA (quiescent or active)), or as continuous values. The first step is defining metrics that are easy to measure in the system, such as the total percent of nodes that are live and providing credit. More complicated models could measure the magnitude and change in magnitude, with quantities at the smoothest part of the credit curve. Multi-level systems analogous to options trading could be implemented, for the relevant metrics equivalent to the price-delta-gamma-vega partial derivative stack in options trading (and also introducing theta as a time parameter to build in temporal dynamics). The metrics could be read either as an interpretation of mathematically smoothness, or as a system-abstracted value such as temperature.

**State Transitions**
The objective of a smart network field theory is to characterize, monitor, and control smart network systems. Therefore, an important goal of a field-theoretic formulation is to be able to identify system phase transitions, both positive and negative, and first and second-order (a change per a continuous parameter or a dynamical evolution of the system).[43] State transitions are the threshold levels of node actions (system-level quantities measured). The point is to understand what happens when the system reaches certain threshold levels of system-wide quantities. The quantities could be measured with temperature terms, operators (Hamiltonian and Lagrangian terms), or variational parameters and system path integrals (action). The idea is to capture the system-wide calculation of the probability that certain nodes are in certain states, and thus that certain network resources are available.

The Hamiltonian-type term in the smart network field theory could measure system quantity metrics as system-wide resource indicators, and economic indicators as a class of risk-assessment metrics. In a practical example, a smart network field theory could be used to understand and evaluate systemic risk and the possibility of flash crashes. This is particularly important now that trading is a multi-species activity, conducted half by human agents and half by programmatic trading in worldwide equities markets.[11]

Not only is a temperature term a useful macroscopic control mechanism for a system comprised of a huge volume of microscopic activity, it is necessary in systems in which microscopic-level data may not be available. Mechanisms such as smart network field theories could be important in domains such as smart networks for network resource monitoring and control because less information is available. In smart networks, less microscopic information may be available, so what is needed is a meta-information overlay regarding the aggregate activity of the system. Some of the ways that less information is available in the privacy-protected computing era of smart networks is 1) confidential transactions (using Monero, Z-cash, and Ethereum), 2) permissioned enterprise blockchains (limited transaction visibility to participating parties), and 3) Data Markets (unlocking new data silos (health information, validated social network referrals) with private transactions). The question is how to obtain an overlay metric that provides aggregate information about the computing network. Similarly, in deep learning, the architecture is by definition having hidden layers whose micro-computation is unknown at the macroscopic level. This is an important result of smart network field theory. The smart network architectures require new models for their evaluation, which smart network field theory provides.

A defining feature of smart networks is that they are simultaneously private and transparent, meaning that the whole system operates per the same parameters (transparent), but the individual microscopic level transactions (blockchain) and forward-back-propagations (deep learning) are hidden. Therefore, meta-level temperature terms are needed to understand overall system behavior. Accurately-calculated mathematical formalisms per smart network field theories are a means of deriving a well-formed temperature term for a network, especially because the microscopic-level "particle movement" is a system parameter that is unknown.

**Economic Indicators of the Future**

The amount of available network resources (system-wide quantities) might be calculated as future Economic Indicators. The idea is to use smart network field theory to measure network resource availability, and mange (or have self-managed) economies based on this. Ripple already constitutes a new form of real-time economic indicator in the sense that it is a live credit network. Open credit links create a live network for monetary transfer and are a real-time measure of economic confidence beyond any currently existing metrics. As credit-extending nodes retract their links, the measure of economic confidence and credit availability decreases.[44] The credit elasticity of the Ripple network could be a smart network field-theoretic calculation that includes temporal aspects as live credit varies over time. Likewise, the idea of real-time balance sheets (as all significant worldwide assets inventories might become blockchain-registered) could provide a stock market-like immediate valuation of assets, as opposed to *a posteriori* calculations after the accounting period. A first step could be shifting current economic indicators to real-time calculability with smart network field theory. In the same way that the foldability of a protein is calculated, so too could smart networks detect the rate of new business formation from SME (small and medium-size enterprise) activity.

Another future metric could be a real-time supply chain calculation of the cost to ship a kilogram of good per kilometer of distance worldwide (the "Maersk path integral"). This could be similar to the metric of the cost to lift a kilogram of cargo to space.[45] SpaceX has lowered this by an order of magnitude, currently citing a price of \$4,700/kg for the existing Falcon 9 and \$1,700/kg for the future Falcon Heavy.[46] Defining metrics can be a galvanizing step in focusing an industry on improvement. An example of this is the cost to sequence a single human genome. Using smart network field theory to define and measure metrics could be important for future-class projects. Some of the farther consequences of smart network field theory are using it as a blueprint, as a formal means and planning mechanism, for defining the progress to the realization of higher-level scales of Kardashev society (the ability of civilization to create and deploy advanced technologies), as a real possibility for constructing beyond-terrestrial human futures.

From a smart network field theory perspective, the important point is to measure and predict the amount of network resource availability in the past, present, and future. The network should signal information about resource availability.

**Section 7: Smart Network Service Provisioning and Application Layers**

To concretize smart network field theory, this section outlines some specific examples of the kind of system resources that might be measured by a Hamiltonian-type term, and how system dynamics might signal system criticality and phase transition. Smart network field theories might be used to provision and deploy both network resources and application layers on the network stack. Similar to the operation of communications network provisioning, two of the most straightforward application layers could be basic services and value-added services.

Nodes might provide peer-based services to other peers in the network, whether blockchain p2p nodes or deep learning processing nodes. Figures 11 and 12 list some of these kinds of services for blockchain networks and deep learning networks, stratified by two classes, basic services and value-added services. The most general "business model" in blockchain networks is a peer-based transaction fee for services or other credit assignation mechanism, and in deep learning networks is a loss function optimization.

**Figure 11**. Blockchain Network Services provided by p2p Nodes.[44]

| Network Services | Type of Service provided by p2p Nodes | Specific Solution(s) |
|---|---|---|
| Basic Services | Transaction execution, network security, wallet identity management (public-private key pairs) | Transaction confirmation and network security |
| | Record-keeping, archival | Ledger hosting |
| | Data storage | Storj, IPFS/Filecoin |
| | Software updates | Core developers, improvement proposals |
| | Routing (blockchain routing protocols are open, just as internet routing protocols (TC/PIP) are) | Custom routing services based on cost, speed, confidentiality |
| Value-added Services | News, Social Networking | Steemit, Yours |
| | Intellectual Property Registration | Proof of Existence, Monograph, Verisart, Crypto-Copyright |
| | Banking and Finance | Payment channels (Lightning, Plasma), credit (Ripple) |
| | Digital Identity Management, Reputation | Evernym, uPort |
| *Token Services* | Local economy: community energy-sharing | Transactive Grid (Brooklyn NY), ION |
| *Token Services* | Local economy: voting, community management | District0x, DAOdemocracy |
| | Transportation | Lazooz |
| | Education certification (diplomas, transcripts) | MIT Digital Certs |

Basic services include administrative services that are expected for the orderly and secure operation of the distributed system. These services include transaction confirmation, network security, mining and consensus algorithms, record-keeping (ledger-hosting), and identity management (addresses, public-private key pairs). Another tier of services, value-added services, run as a network overlay. Value-added services might include peer-based hosting of news and social networking applications, intellectual property registration (with hashing and time-date stamping), and digital identity management and reputation. Banking and financial services could be provided via payment channels, and credit (open credit links on the Ripple network). Digital asset inventories (registration, tracking, pledging, and contracting) could be another service. Community participation and orchestration is another value-added network service. Services could include token issuance and management, resource access, voting, and supply-demand matching. An example of local economy management services could be having an Uber-type app to coordinate the local peer-based economy for microgreen harvesting.

**Figure 12**. Deep Learning Services provided by Perceptron Nodes.[47]

| Class of Service | Type of Service provided by Perceptron Nodes | Applications |
|---|---|---|
| Basic Services | Data classification (train on existing data) | Facial recognition, language translation, natural language processing, speech-to-text, sentiment analysis, handwriting recognition |
| | Feature identification, multiple levels of abstraction | |
| | Data identification (correctly identify test data) | |
| Value-added Services | Pattern recognition | Medical Diagnostics, Autonomous Driving |
| | Optimization, error correction | Supply Chain Analytics |
| | Time-series forecasting, prediction, simulation | Error contribution |
| | Data automation | Privacy-Protected Data Markets |
| | Memory function, sequential data processing | LSTM RNNs |
| *Advanced* | System architecture design and improvement | Dark Knowledge |
| *Advanced* | Advanced learning techniques | Adversarial Nets, Reinforcement Learning |

**Smart Network: Basic Administrative Services**

Generalizing from the blockchain and deep learning examples, a smart network field theory could help aggregate and manage a variety of administrative activities. This is the basic class of administrative services that any smart network would be expected to have. Some of these administrative activities include security, activity logging, and system maintenance. Within security, there could be identity confirmation, antivirus services, and overall network security. Within activity logging, there could be operational execution, secure audit-logging, backup, lookup and information retrieval, search, and data aggregation. Within system maintenance, there could be software updates, hardware scans, and other kinds of updating.

**Smart Network: Value-added Services**

At another level, there is a class of applications enabled by smart network field theory that facilitates the practical engagement of the network to develop and provide more complicated value-added layers of activity. This level of activity enrichens and makes the network and its activities more substantive. The perceptron Hamiltonian for example, could signal factors of novel emergence that could then be translated into new value-added services to be incorporated into the deep learning network.

*Systemic Risk Management*

An important class of value-added services is risk management. Systemic risk has been shown to increase as a function of the coupling strength between nodes,[48] so the suggested risk-management service would be using smart network field theory to obtain a global measure of coupling strength between nodes, and understand how this might directionally impact systemic risk.[49] A problem that might be helped with smart network field theory is financial instability in the form of flash crashes, which occur at faster-than-human-manageable time scales.[50] Some of the proposed solutions for managing flash crashes include using statistical models for phase transitions and Jones polynomials (which model the stock market with a knot and braid topology).[51] The idea would be to take these methods and apply field-theoretic principles such as attempting to avoid local minima in a random energy landscape, to likewise avoid the analogous singularity in a smart network system, behavior such as double-pendulum chaoticity that could result in flash crashes.[1,2]

*Collective Intelligence as an Emergent System Property*

Collective intelligence might be harnessed as an emergent system property, and provisioned across the network. Deep learning chains could be used to provide pattern recognition and privacy-protected computing services such that agents could have validation about the aggregated use of their information. One example could be a *blockchain Hamiltonian* for market sentiment. Another means of harnessing collective intelligence could be via the self-play method, as are used now between human agents, and in video games and adversarial deep nets. To reduce systemic risk, trade bots could be pitted against one another in a public goods game in which the (remunerated) goal would be avoiding system criticality such as flash crashes. Human traders and investment houses could pay a small insurance fee per trade (commission) into the incentive pool for programmatic trading entities to prevent financial contagion. Collective intelligence could become a provisionable network resource (Figure 13) for systemic risk management. The benefit of such a CrashWatch metric could be earlier warning signals in the case of market crashes, bringing more resolution to their effective management long before system criticality (a crash) happens. In another smart network system, UAVs, collective intelligence is being harnessed as an emergent system property. A decentralized air traffic navigation model based on programmed flocking and collective behavior arising through information sharing has been demonstrated.[10]

*Network Resource Provisioning*

Another important class of value-added services is resource discovery and production. Field theories could help by detecting emergence in the network, not only for system criticality and risk management purposes, but also for new resource identification. Smart network field theory might be employed to find, measure, create, and deploy network resources. First, it is necessary to locate, measure, and deploy existing network resources. Second, it is necessary to identify emerging network resources and new layers (and types of layers) in the network stack. Third, for the generation of new resources, field-theoretic principles could be used to design, test, and simulate new resources before they are widely deployed to the network. For example, as smart network nodes (deep learning perceptrons and blockchain p2p nodes) might become more autonomous, the idea would be to canvas the nodes to gauge the network demand for deploying the new resources and previewing what new kinds of services might be delivered with the resources. Having more proactive and real-time resource-demand forecasting is one way that *supply chains* could shift to becoming *demand chains* instead. One such "demand chain application" could be predicting demand from user attributes and aspirations which are shared in privacy-protected Data Markets. These kinds of next-generation social networking applications have been articulated as value-added overlays to the social network structure.[52]

**Figure 13.** The Smart Network Stack for Resource Provisioning.

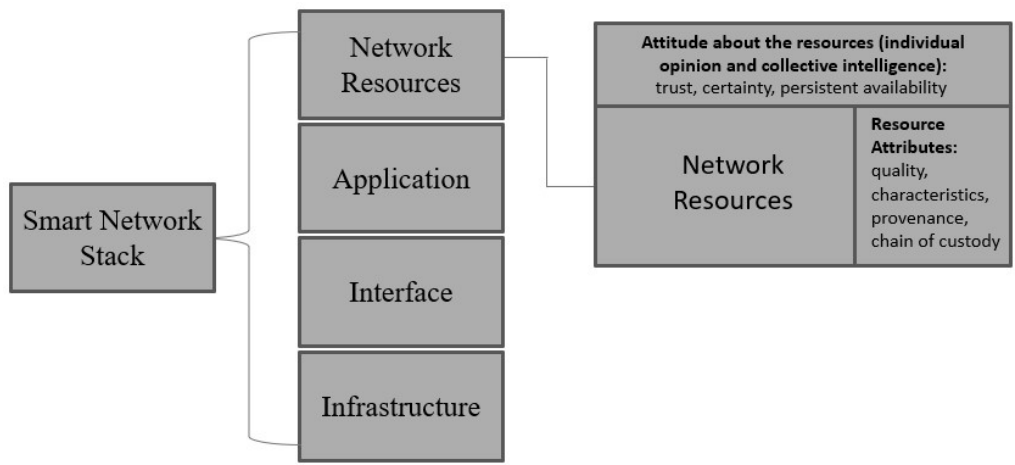

The Smart Network Stack is highlighted in Figure 13, with the four main layers of infrastructure, interface, application, and network resources. A smart network field-theoretic function such as a Boltzmann ensemble distribution (a probability distribution over all possible states of the system) could be used to measure the availability of network resources across layers. Just as synaptic excitation is a network resource that triggers the action (the motion of the overall network), so too are smart network resources such as algorithmic trust, information attributes, and liquidity (for example, automatic transaction financing and supply-demand matching). Another smart network field-theoretic function could be used to mobilize resources as the *propagation of certainty*. The idea is to take the inverse of Gibbs' *propagation of uncertainty* as a quantity diffusing through the network that is the *propagation of certainty* which sums, measures, and calculates the quantity of available network resources, such as algorithmic trust. Certainty is a meta-resource, an information layer about the surety of the availability of resources, which targets the same need as a futures contract (guarantee of future resource availability) with a smart network market mechanism.

## Section 8: Implications of Smart Network Field Theory for Concurrency Computational Complexity

One implication of smart network field theory is that it is a complexity method that can be applied to the consideration of computational complexity, and possibly offer alternative ways of casting the P/NP problem schema. The structure for the classification of computability problems is the P/NP (P, NP, NP-Complete, or NP-Hard, and beyond) framework that organizes problems into different classes based on the time and space needed to solve them, including many delineations such as polynomial (regular) time, exponential time, and double-exponential time. More generally, *computational complexity* is the classification of computational problems based on the difficulty of computing them, as constrained by the space and time resources in computing systems. The premise of smart network systems is that they are fundamentally a new form of distributed global computational architecture, which is no longer subject to the traditional time and space constraints which structure the P/NP schema, which suggests recasting the problem schema based on other parameters. For example, what if the schema were not organized based on constraint, but on possibility? What might be solved if computing resources were not a barrier? Advance in rethinking computational complexity has been obtained from the consideration of quantum gravity[53] and black holes[54] as alternative kinds of information domains, and the advent of smart networks might sponsor the same.

The P/NP framework for evaluating computational complexity might be rethought in the context of smart networks because p2p transaction networks and deep learning hidden-layer networks might constitute novel computational domains. One vector to explore is the network architecture itself for advanced modes of scientific computation (which has been researched in the form of p2p desktop grid computing[55]). Another vector to examine is lifting classical space and time limits. Smart network computing may not have the same constraints of the traditional methods of evaluating P/NP-hard problems limited to running on von Neumann architectures and parallelized supercomputing architectures. The native time and space domains of smart networks are distinct: temporally with blocktime and learning network solve-time; and spatially with arbitrarily-expandable distributed p2p nodes, hidden processing layers, and self-optimizing architectures. New means of structuring computational problems could trigger an alternative orientation to computational complexity, and a respecification of the P/NP schema and lead to advance in computational concurrency.

### *Smart Network Implications for Concurrency and Computational Complexity*
Smart networks are non-linear systems in which there are multiple time and space scales. Therefore, a field theory is needed to calculate the system state and criticality, with macroscopic temperature terms, across diverse temporal and spatial regimes. The field theory is a method to align smart network systems with computation concurrency principles given the different (and dynamical) time and space scales. Formally, smart network field theories accomplish this concurrency management with a renormalized action or energy operator (Hamiltonian or Lagrangian) corresponding to an overall system state. Also (a temperature term or variation term). Computability as a basic feature is also built into the smart network field theories. For example, one of the *temperature terms* in the Wolynes Spin Glass Model is a quantitative measure of the computability of a protein: how much memory and time would be needed for the computation, and whether the problem is computable.

The smart network system manages concurrency with complexity methods, which thus introduces parameters of time and space malleability. Smart network field theory creates a method for working with space and time as variable parameters. The ability to engage space and time as malleable system parameters is a formal method, constitutes not just the concept, but the mathematics for recasting the P-NP computational complexity problem schema. Smart network field theory could be used as an alternative framework for analyzing P/NP computational complexity. Smart networks may not be subject to the same time and space constraints of classical computational resources, and thus invite innovating thinking about the conceptualization of computational complexity, and the structuring of computable problems.

### *Smart Network Concurrency Application:*
### *Lateral-Medial Coordination of Human-Machine Musculature*
To concretize the concurrency feature of smart network field theory, a specific application can be articulated. The problem domain is the lack of concurrency (overlap and coordination) between human and robotic muscles, which need to be orchestrated and aligned for automation in contexts such as factory work, physical therapy, ergonomic design, and medical devices. The issue is that the muscles of living organisms tend to be laterally-oriented (toward arms and legs, away from the midline plane), whereas machine-designed robotic articulation are typically medially-oriented (toward the midline, or medial plane).[56] The concurrency of these two different domains could be orchestrated with a smart network field theory to help manage the complexity of ongoing automated mapping between human and machine systems as partners in dynamical interaction. A smart network concurrency technology is needed: deep learning chains for the secure automation, audit-log, and liability-tracking of the effort (blockchain), together with data collection, object identification, and pattern recognition (deep learning).

### Computational Advance
A key requirement of a field theory is extensibility to uncertain future situations, especially in a domain such as smart networks, which is itself new and evolving. Any concurrency control theory needs to include the consideration that smart networks are not only a new class of global computational infrastructure, but also one that is itself leading the vanguard of computational infrastructure development, such as tensor processing, automatic value transfer, and quantum computing. The expectation would be that these two areas may continue to evolve in lock-step, emerging computational paradigms, and smart network technologies such as blockchain and deep learning, since one is driving the other.

### *Quantum Computing*
Quantum computing simultaneously poses the biggest immediate threat and opportunity to blockchain as a smart network technology. First, although quantum cryptography might be able to break the current cryptography used by blockchains (mainly SHA-256), blockchains might likely upgrade to quantum cryptography as methods become available, and there could be quantum-secured blockchains.[57] Second, since quantum cryptography would only be adding a quantum layer to the standard blockchain protocol, next-generation projects might more fundamentally change computing architectures and instantiate the blockchain itself with quantum computation. One proposal is to create a blockchain using quantum particles that are entangled in time (encoding the blockchain into a temporal GHZ (Greenberger-Horne-Zeilinger) state of photons).[58] The idea is that data would be encoded on a quantum particle, which would store the history of all its predecessors in a way that could not be hacked without destroying it.

Quantum computing is also already being considered for use in deep learning networks, particularly to reduce processing time and resource utilization.[59] Google, through its Quantum Artificial Intelligence Lab, launched in 2013, continues to explore quantum computing technologies particularly through hybrid classical-quantum systems, in specific applications such as quantum simulation, quantum-assisted optimization, and quantum sampling.[60] Recent work focuses on chemistry, quantum annealing, and classification.[61]

The point is that research programs in smart network technology areas (blockchain and deep learning), have been making significant investments in quantum computing. The benefit of having a field theory that is already expressed in quantum terms (with statistical mechanics and information theory), is possible extensibility in transitioning from classical computational paradigms to quantum computational paradigms.

### Smart Network Challenges

Smart network concurrency methods might help address some of the key challenges facing smart networks. In blockchain and deep learning, these challenges include blockchain scalability (growth and efficiency), and extending deep learning beyond simple object recognition to more complicated applications (such as time-series forecasting). Smart networks face challenges in that the large volume of microscopic-level behavior is unwieldy (blockchain) and inextensible (deep learning). The first blockchains, Bitcoin and Ethereum, may not be scalable from their current 7 transactions/minute processing to Visa-class processing (2,000 transactions/second), even with proposed initiatives such as side chains and payment channels (contractual structures for ongoing transactions). Likewise the core self-operating feature, consensus algorithms (proof of work and proof of stake may require next-generation PBFT algorithms). Likewise, in deep learning, it is not clear how to extend success in basic object recognition to other more complex domains. For example, Google's AlphaGo system directed at the task of Atari game play (47 classic arcade games) found that while the system can learn and beat human competitors at any particular game, the learning is not portable.[62] The system can only play one game at a time, and must learn the rules from scratch for each game. The Smart Network Field Theory can help by packetizing and quantizing system microscopic-level behavior to define "temperature terms" at the macroscopic level.

### Risks and Limitations

There could be many risks and limitations to the smart network field theory development proposed here. One class of risks concerns the theory development. A charge of over-reaching could be levied, that the ideas are too much of an extrapolation. It is too early to propose such a field theory, the underlying equations are not included, and specific data applied to a use case for a demonstration proof is also not provided. Not only are the formalisms not yet worked out, it is unclear if and what data would be available to test this theory, especially for the very-large data sets that would be required. Further, the underlying smart network domains may be too heterogeneous for the general conceptual notion of a smart network field theory to be useful. However, given the fast pace of adoption of smart network technologies such as blockchain and deep learning, the counter-argument would be that it is not too early to begin the formal theorization of this domain.

Another class of risks concerns the application of the theory. A critique could be raised that even if applied, smart network field theories would not necessarily guarantee any kind of problem resolution. The phase transition disasters (such as financial crashes) that smart network field theories might purport to be able to identify and manage could fail to be obtained as a practical result. Phase transitions are an aspect of complex systems that perhaps cannot be predicted ahead of time, or with enough time for intervention. The kind of counter-action, and the degree to which such a response might be helpful is not clarified (and is unknown). A model is only a representation of the territory it attempts to map. Aspects may arise in the underlying system that are outside of the model's frame. However, exactly because smart networks are complex systems that might self-tune their behavior in unexpected ways, theoretically-based methods for studying them and trying to manage them could be useful. The methods proposed by smart network field theory, are at minimum, not futile, and could have the benefit of sponsoring a more rigorous study of complex systems.

## Conclusion
### *Overview of Results*
This paper defines *smart network field theory* as a conceptual and formal tool for the characterization, monitoring, and control of smart networks. *Smart networks* are a new form of global computational infrastructure such as blockchain economic networks and deep learning pattern recognition networks, and may also include autonomous-strike UAVs, smart energy grids, algorithmic trading, and the real-time bidding market for advertising. Smart networks are complex systems, and therefore, a well-grounded systems-level theory, with robust foundations in physical science is needed. Thus, this work proposes *smart network field theory*, developed from statistical physics, effective field theories, and model systems (Cowan's Statistical Neural Field Theory and Wolynes's Spin Glass Model), for the purpose of criticality detection and fleet-many item orchestration in smart network systems.

Smart network field theory is articulated as having three minimal constituent elements. These include a temperature term, an operator (Hamiltonian, Lagrangian) or variational principle (action, path integral), and portability across system scales (discussed in Section 5). The first element is a temperature term. Of particular interest in complex systems is identifying *temperature terms* that emerge at the macroscopic level and correspond to a large volume of activity at the microscopic level. Some examples of the temperature term in smart networks are hash rate (blockchains) and error rate (deep learning nets). The temperature term provides a macroscopic label of an informational state of the system.

The second element is an operator or variational principle such as a *Hamiltonian term*, which is a point value that captures an entire configuration of the underlying elements in a dynamical system. In smart network field theory, one way of instantiating the Hamiltonian term could be as a formalized measure of *system capacity*. The "blockchain Hamiltonian" or "perceptron path integral" might identify the number or percent of system nodes that are available to provide services such as credit availability (blockchain) or activation (deep learning) at a given time.

The third element is portability across system scales, which is the ability to span multiple levels within a system. Smart network technologies such as blockchain and deep learning are not only one microscopic and macroscopic level, but may extend many levels up and down, quantitatively and qualitatively. Merkle roots (blockchain) and long short term memory (LSTM) (deep learning) are two system-spanning mechanisms that have evolved within these smart network systems. Since smart network systems are multi-tier, the Hamiltonian term might be used to identify the appropriate system level for certain monitoring or intervention (taking the partial derivatives of the Hamiltonian term as a means of determining where the curve is the smoothest).

### Farther Consequences of Smart Network Field Theory

The biggest potential practical benefit of *smart network field theory* is that it might be used to understand and manage smart network systems as they continue to evolve. Another potential benefit is using smart network field theories as the basis for designing, testing, and deploying new smart network systems, and for provisioning additional network layers and resources within existing smart network systems. Smart network field theories could lead to the well-formed engineering and instantiation of complex systems. A farther practical benefit of smart network field theories is that they might be used to identify novelty such as system criticality and phase transition, for risk management (for example, understanding and avoiding flash crashes), and for beneficial emergence (for example, articulating new systems-level resources such as algorithmic trust (blockchain) or time-series forecasting capability (deep learning)). Another practical benefit of this work is the informational resource created by the detailed summary of the model systems (Cowan's Statistical Neural Field Theory and Wolynes's Spin Glass Model), highlighted in Section 4, and theoretically derived in the Appendices.

Some of the farther theoretical consequences of this work are two-fold. First, the work suggests the possibility of recasting the P/NP computational complexity schema as one no longer based on traditional time (concurrency) and space constraints, due to the availability and novel architectures of smart network computational resources as fundamentally a new form of global computational infrastructure. The field theory is a time-space malleability method for managing system concurrency and simultaneity. In addition, large-scale network analysis methods such as smart network field theory, with foundations in statistical physics, information theory, and complexity could help to more firmly establish information theory as a 21st century paradigm for the conduct of science.

Second, articulating the notion of smart network field theories as a means of assessing system criticality provides a foundation for the further study of this topic. The technophysics concern of system criticality could become more of a formal area of study, as complexity theory has, with theoretical constructs and mathematical formalisms. System criticality and phase transition is of interest, but precursors for its avoidance are also important, especially from a complex systems management perspective. Smart network field theories might be helpful in eliciting interim system structure between the levels of microscopic noise and macroscopic labels.

*Word count (without Appendices): 15,213*

**Glossary**

**Adversarial networks:** Adversarial networks is a self-play method used in deep learning in which an adversary network generates false data and a discriminator network tries to learn whether the data are false

**Blockchain (distributed ledger) Technology:** A blockchain is an immutable, cryptographic (cryptography-based), distributed (peer-based), consensus-driven ledger. Blockchain (distributed ledger) technology is a software protocol for the instantaneous transfer of money and other forms of value (assets, contracts, public records, program states) globally via the internet. Other topologies of distributed ledger technology may include Directed Acyclic Graphs (DAGs), in projects such as IOTA, Hashgraph, Byteball, and DAGCoin.

**Cloudmind:** A Cloudmind is a cloud-based collaboration of human and machine minds (with safeguards and permissions). 'Mind' is generally denoting an entity with processing capacity.

**Cryptoeconomics:** Cryptoeconomics is an economic transaction paradigm based on cryptography; more specifically, an economic transaction system implemented in a cryptography-based software network, using cryptographic hashes (computational proof mechanisms) as a means of confirming and transferring monetary balances, assets, smart contracts, or other system states. A key concept is *trustless* trust, meaning relocating human-based trust to cryptography-based trust

**Dark Knowledge:** Dark knowledge is a technique used in deep learning in which the predictions of several model runs are compressed into a single model to generate more efficient results

**Deep Learning Chains:** Deep learning chains are a smart network convergence technology that includes the functionality of blockchains and deep learning. Deep learning chains allow the secure trackable remunerated automation of fleet-many items (blockchain) together with object identification, pattern recognition, and optimization (deep learning) for applications such as food traceability, autonomous driving, space mining, and medical nanorobot operation.

**Deep Learning Neural Networks:** Deep learning neural networks are computer programs that can identify what an object is; more technically, deep learning is a branch of machine learning based on a set of algorithms that attempts to model high-level abstractions in data by using artificial neural network architectures, based on learning multiple levels of representation or abstraction, such that predictive guesses can be made about new data.

**Field:** A field is 1) the precise physical definition of an electromagnetic or gravitational field; 2) the ability to control fleet-many items as one unit; 3) the situation that every point in a landscape has a value which may be calculated (per effective field theories in physics)

**Hamiltonian term:** A Hamiltonian term is an operator (Hamiltonian, Lagrangian) or variational principle (action, path integral) which is a point value that captures an entire configuration of the underlying elements in a dynamical system, for example as a formalized measure of *system* *capacity* (the number or percent of system nodes that are available to provide services such as credit availability (blockchain) or activation (deep learning) at a given time)

**IDtech (identification technology):** IDtech is the (increasingly assumed) ability of technology to recognize objects; similar to FinTech, RegTech, TradeTech, and HealthTech; technologies that digitize, standardize, modularize, and automate their respective domains

**P-spherical Spin Glass:** A p-spherical spin glass is a spin glass model in which the spin-up and spin-down values are numbers corresponding to the spherical coordinates of a point P

**Smart Contract:** A smart contract is a software program registered to a blockchain for confirmation (time-datestamping provenance), and possibly automated execution in the future. To be legally-binding, smart contracts need to have the four requisite elements of contracts: two parties, consideration, and terms

**Smart Networks:** Smart networks are intelligent autonomously-operating networks

**Smart Network Field Theory (SNFT):** A *smart network field theory* is an effective field theory or any formal method for the characterization, monitoring, control of *smart network* systems such as blockchains and deep learning networks, particularly criticality detection and fleet-many item orchestration. Special forms of the SNFT for very-small and very-large systems could include *Biological Smart Network Field Theories (bSNFT)* and *Space-based Smart Network Field Theories (sSNFT)*

**Smart Network Technologies:** Exemplar smart network technologies are blockchain economic networks and deep learning pattern recognition neural networks

**Spin Glass:** A spin glass is a disordered magnet that is a metastable system with half of its molecular bonds oriented in the direction of spin-up, and the other half in the direction of spin-down. The term *glass* is meant as an analogy between the magnetic disorder in the spin glass and the positional disorder in a conventional glass. The benefit of using a spin glass model is that the spins can be relaxed into real numbers, so that systems can be solved analytically

**System scale portability:** System scale portability is the ability to span multiple levels within a system, for example using Merkle roots (blockchain) and long short term memory (LSTM) (deep learning). Taking derivatives of the Hamiltonian term might indicate optimal points for system intervention (where the curve is the smoothest)

**Technophysics**: Technophysics is the application of physics to the study of technology; specifically, the study of technology with physics-based models, concepts, and methods such as statistical physics, Brownian motion, the Schrödinger equation, Shannon entropy, information theory, and computational complexity; by analogy to biophysics and econophysics

**Temperature term:** A temperature term is a consolidated measure that comprises all of the activity of a microscopic level at one or more tiers down in a system. The temperature term provides an aggregate of an informational state of the system and might be employed as a control lever for the system.

# Appendix 1
## Model Field Theory #1: Statistical Neural Field Theory (Cowan)

**Figure A1**. Major Steps in the Derivation of Cowan Statistical Neural Field Theory.

| | Problem/Requirement | Tools/Techniques | Results |
|---|---|---|---|
| A | Description of the system at equilibrium | • Mean Field Theory<br>• Continuum approximation | • Wilson-Cowan equations (a mean field theory) |
| B | Description of the system at criticality | • Effective Field Theory | • Cowan Statistical Neural Field Theory (a statistical field theory)<br>• A statistical field theory is necessary to describe the effects of correlations and fluctuations in dynamical non-equilibrium systems |
| 1a | Write a system master equation considering all possible state transitions | • Brownian motion, diffusion limit, path integral | • System master equation written in the form of a Markov random walk |
| 1b | Formulate the system as a Markov process | • Path integrals, Green's function (Gaussian propagator), Wick's theorem<br>• Take the diffusion limit of the Markov random field and derive path integrals | • Path integrals derived: correlation functions, moment-generating function (an alternative probability distribution for the system), and a system action<br>• The action (the path integral of the Lagrangian (the overall configuration of states)) is used to construct the moment-generating function (alternative probability distribution) that describes the statistics of the network |
| 2a | Generalize a single random walk to random walks in a continuum to expand from Markov random walk to Markov random field | • Quantize neurons into three states (QAR) | • Moment-generating Function is rewritten for Markov random fields<br>• Generalize the action over the network with Green's function (path integral)<br>• Master Equation rewritten as Neural Master Equation, comprising all system states, for the Markov random field |
| 2b | A model of state transitions | • Gell-Mann 3x3 matrices, ladder operators, algebraic simplification | • Algebraic simplification of Master Equation and Solution of the Master Equation expressed as a path integral of the Markov random field<br>• Number density operator to count: 1) the number of neurons in the QAR states, 2) total network excitation<br>• State diagram of the network (with neural state vector, probability state vector, neural Hamiltonian (counting all possible state transitions)) |
| 3a | A linear model of the system action | • Wiener-Feynman Path Integral interpreted in Euclidean QFT:<br>• Simplified Spike Hamiltonian<br>• Coherent States | • Rewrite Moment-generating Function with Coherent States (obtain Coherent State Path Integral)<br>• Linearized Action in the form of a Linearized Spiking Model |
| 3b | A non-linear model of the system action (firing rates are non-linear across the neural field in reality) | • Standard simple perturbation (Renormalization Group) | • Non-linearized Action in the form of a Renormalized Action<br>• System dynamics in the form of a reaction-diffusion system to be used in determining System Criticality |

| 4a | A model to assess System Criticality | • Reggeon Field Theory | • Renormalized Action = a Reggeon Field Theory Action<br>• Reggeon Field Theory systems are those with DP Phase Transitions |
|---|---|---|---|
| 4b | A model to detect Phase Transitions | • Directed Percolation Phase Transition | • A non-equilibrium phase transition model that adequately describes System Criticality: Ability to describe and measure system criticality |
| 5 | Mechanism to predict and control System Criticality | • Optimal Control Theory<br>• Synaptic Plasticity Operator | • Ability to predict and manage system criticality (in applications)<br>• Shorter timescales to produce or avoid self-tuned system criticality<br>• Path integral of the OCT Action |

## Model Field Theory #1: Statistical Neural Field Theory (Cowan)

## Methods

This section sets forth Jack Cowan's Neural Statistical Field Theory as prepared from his research corpus as a mathematical neuroscience researcher at the University of Chicago. Specific resources include Cowan's complete account of developing neural field theory (2014),[63] Cowan's publications and lectures,[64] Destexhe and Sejnowski's account of Wilson-Cowan 36 years later (2009),[65] Bressloff et al.'s "Festschrift for Jack Cowan" paper (2016),[66] and recent work applying deep learning techniques to the modeling of the visual cortex.[67] The aim of this section is to provide a summarized description of the themes and concepts in the field theory in accessible terms for non-practitioners.

## Introduction: Neural Statistical Field Theory

Cowan develops a field theory of large-scale brain activity using statistical mechanics. For the basic network description in stable non-critical situations, a mean field theory called the Wilson-Cowan equations is articulated (1972).[68] To assess system criticality and phase transitions, a more complicated field theory is proposed in which network behavior evolves over time, and is represented with Markov random walks and Markov random fields as a path integral (2007).[69]

## A. Mean Field Theory to describe Stable Equilibrium Systems

The Wilson-Cowan equations are a continuum approximation (a means of modeling system kinematics as a continuous mass) inspired by work from Cragg and Temperley (1954) which examined ferromagnetic atomic interactions in physics applied to neurons.[70] Whereas Cowan applied anti-symmetric weights to construct a theory of coupled oscillators (excitatory and inhibitory oscillators), Jack Hopfield applied the Wilson-Cowan equations with symmetric weights, thereby inaugurating the development of the field of machine learning. Hopfield's intuition was that an Ising model of ferromagnetism could be used for modeling neurons.

There could be an analogy with a spin glass (a disordered magnet that is a metastable system with half of its bonds spin-up, half spin-down), using spin glasses as a physical system in which energy can be minimized as a computational feature. The Hopfield network was thus developed using these principles to create an artificial neural net (modeled on a computational model of brain neurons) in 1982.[71] A related model, Boltzmann machines, was proposed in 1983,[72] in which an "energy" term is also defined as an overall network parameter. The sigmoidal structure (i.e. a logistic regression formulation of the underlying problem) proposed by Cowan also made a significant contribution to machine learning. The sigmoid is a key structure for solving the credit-assignment problem for perceptrons (allowing an algorithm to easily interpret an s-curve as yes-no binary values or probability percentage values).

## B. Statistical Neural Field Theory to describe System Criticality and Phase Transition

The mean field theory articulated with Wilson-Cowan equations does a good job of describing stable systems that are at equilibrium. However, to articulate the effects of correlations and fluctuations over time in a dynamical non-equilibrium system, a more robust model is needed, such as a statistical field theory, which captures the state transitions and the potential movement of the system.[73] Cowan was inspired by Per Bak's work on self-organized criticality (1987),[74] and Softky and Koch's documentation of anomalously-large fluctuations in spontaneous neural spiking patterns (1993).[75] Summarizing the overall method, to assess criticality and the effects of correlations and fluctuations in the non-equilibrium system, Cowan sets up the evolution of neural activity as a Markov process, and takes the thermodynamic limit or continuum limit of that expression. The continuum formulation means ending up with a Markov random field which can be manipulated with path integrals and other calculus-based tools, as described below.

### 1a. Master Evolution Equation for a Markov Random Walk

The first step is to obtain a random walk, by writing a master equation with neurons on a d-dimensional lattice (hypercube). The master equation is an evolution equation that considers all possible state transitions of the system. Structured as an evolution equation, the diffusion limit can be taken, with the spacing between the neurons going to zero, and the firing time going to zero. The diffusion approximation of the random walk is obtained (diffusion is relevant as the physical observation of the Brownian motion of molecules in a random walk).

Norbert Wiener (1958)[76] introduced the mathematical ideas of Brownian motion and the diffusion limit (although Einstein (1905)[77] was first to study these problems in physics, and Bachelier in the context of finance (1900)[78]). The solution of the diffusion equation is a known quantity: a Gaussian propagator or a Green's function (a path integral that can be used to solve a whole class of differential equations). Wiener expressed the solution of the diffusion equation as a path integral, from which the action (an abstract quantity describing the overall motion of a system) can be derived. The path integral is structurally similar to a Hamiltonian (the total energy in a system). Hence, the same kind of mathematics is used here to express the path integral as an overall measure of the system from random walks.

### 1b. Path Integrals for Correlation Functions and the Moment-generating Function

Importantly, correlation functions (measures of distance in the network) can be expressed as the sum over all paths of the system, as path integrals. A moment-generating function can be introduced as the path integral associated with the system. A moment-generating function is an alternative statistical means of specifying a probability distribution (in contrast to probability density functions and cumulative distribution functions), based on computing the moments in the distribution. Cumulants (another set of quantities which provide an alternative to the moments of the distribution) can be obtained the same way. These processes are well-known. The correlation functions that influence system criticality are expressed in a standard way, through a moment-generating function (that is a path integral).[79]

In Gaussian Markov random walks, the Gaussian propagator or Green's function solution to the differential equations can be used to collapse the function into a two-point correlation function. All of the higher moments depend on the first two moments, so the function can be collapsed under this solution method. For example, the two-endpoint moment function can be used to evaluate only the even moments, so the odd moments vanish, while preserving the general structure of the function. This is a result of Wick's theorem (a method of reducing high-order derivatives to a more-easily solvable combinatorics problem). With Wick's theorem, it is only necessary to compute two-point correlation functions (assuming the one-point function is zero). The result is that a usable action is obtained from the two-point correlation function, which is equal to a Gaussian propagator or the Green's function. Wick also showed that if imaginary time is used in the Schrödinger equation, it becomes the diffusion equation, further underlining the connection between Brownian motion and Quantum Mechanics.

## 2. Expanding the Markov Random Walk to a Markov Random Field
### 2a. Generalizing the Random Walk to a Random Field to Specify a Neural Dynamics
The next step is generalizing from a single random walk to random walks in a continuum, to arrive at Markov random fields. A moment-generating function for Markov random fields can be written. This generates a field that varies over the entire network, and the action generalized from that is a derivative that includes both time and space coordinates. Gaussian random walks and fields were described previously in terms of their associated Green's functions. The same structure can be used to model and propose a **neural dynamics**. The neuron is quantized into three states: quiescent, active, and refractory (QAR). This is a three-state model, but it could just as easily be two (QA: quiescent and active neurons). The kinetics of the model are specified as the transitions between the different states (QAR) with rate functions and thresholds. A **neural master equation** for the Markov random field (not just the Markov random walk as before) is introduced to keep track of the states and transitions in the network.

### 2b. Algebraic Simplification of the Master Field Equation
A standard mathematical basis, 3-by-3 matrices, can be used to model the transitions between the QAR states. Murray Gell-Mann's matrices for the representation of the Lie group SU(3) are a good model for this. The algebra of quarks can be applied to the algebra of neurons. Raising and lowering operators (ladder operators) are obtained from the Gell-Mann matrices. The ladder operators can be used to increase or decrease the eigenvalues of other operators. An algebraic version of the master equation can be written with ladder operators and eigenfunctions (solutions to the set of differential equations) to give the QAR state transition diagrams of the neural network.

To obtain the **neural state vector**, neurons are organized (as previously) in a d-dimensional lattice. The weighted configurations and their probabilities are summed to obtain a probability state vector, and the probabilities sum to one. The parallels between Quantum Mechanics and Neuron Mechanics can be seen again in that the mathematics is nearly the same. In the neural network normalization, the sum of the probabilities goes to one. In Quantum Mechanics, the sum of the square of the modulus of the functions goes to one, and that is the wave equation.

The algebraic equation is a number density operator that can be used to count the number of neurons in the different QAR states in the network. Both the number of active neurons can be counted, and the current or total excitation coming onto the network. The total excitation can be counted by weighting the number density of active states by the weighting function for each active neuron. Thus, the master equation can be written as a **neural Hamiltonian**, an evolution operator which counts all possible state transitions. The analog of the second quantized form of the Schrödinger equation in Quantum Field Theory (QFT), viewed as Markov processes for neural networks, can be written. The result is that the solution of the master equation is expressed as a **path integral** of the Markov random field.

### 3a. Wiener-Feynman Path Integral

Following on from Wick's ideas, physicists have recognized that the Feynman path integrals of QFT are Wiener integrals in imaginary time. Therefore, they can be evaluated, because the Wiener integral is mathematically well-defined. It is possible to compute with the Wiener integral (which is just a Gaussian integral), whereas the Feynman path integral is not mathematically well-defined, and difficult to compute. Feynman knew this, that he could calculate with the path integral if he used imaginary time, and this discipline is now called Euclidean (quantum) field theory.[80,81] Euclidean quantum field theory is one modern form in which statistical mechanics is applied (per Ken Wilson[82]). Therefore, Euclidean quantum field theory is applied to neural networks.

### Defining a Reduced Model using Spike Hamiltonian instead of Full Hamiltonian

A reduced model can be created by appeal to solid state physics. A simpler neural Hamiltonian is generated as the spike Hamiltonian instead of the Hamiltonian of all possible state transitions. To do this, a minimal model is considered, one in which spike (activation) rates are low (the probability of a neuron emitting a spike is low). Most neurons are in the quiescent state Q (thus, Q is close to one), which makes the reduced neural Hamiltonian easier to analyze.

### Coherent States

The next step is deriving the moment-generating function for the field theory using coherent states. Coherent states are of interest because they are good representations to use for expressing the statistics of QFT and Markov fields. Coherent states are functions that do not change their state under the action of the evolution of the wave equation (the Schrödinger equation). Schrödinger articulated coherent states in 1926, soon after introducing the wave equation, and they were further developed by Roy Glauber in 1963 in work on quantum optics.[83] There are different kinds of coherent states. Although it might seem that the appropriate coherent states for the neural field theory would be those that express the Lie algebra SU(3) for the reduced model, better coherent states are the more basic ones introduced by Wigner with U(1) symmetry. Using the U(1) symmetry, an accurate theory of spiking action on a neural network can be obtained.

### Obtaining the Neural Action as a Linearized Spiking Model with Coherent States

With the spike Hamiltonian and the coherent states, the moment-generating function can be rewritten. A **coherent state path integral** is obtained for the moment-generating function. The neural action (the overall motion of the system) is obtained in terms of coherent states, meaning terms that are at most quadratic (Gaussian integrals) and can be computed easily. Next, the model is linearized (by finding a linear approximation to assess the stability of equilibrium points). With a linearized model, the Green's function (a path integral) can be written exactly for the **linearized spiking model**. This is the target quantity, the action as a spiking model; as a linearized spiking model that can be evaluated. The linearized network problem structure can then be solved completely in frequency space (Fourier space or momentum space (in physics)) for this network.

### 3b. Obtaining the Neural Action as a Non-linear Renormalized Action

It is possible to use a linearized spiking model to describe the linear model. However, in more realistic neural models, the firing rate function is non-linear across the field, so it is necessary to conduct a power series (Taylor) expansion of the firing norm or probability function. In this case, the action is no longer quadratic, so the perturbation methods of QFT must be used for the calculation. The singular perturbation technique that physicists have worked out is the renormalization group method.[84,85] A renormalization group is a mathematical apparatus that allows systematic investigation of the changes of a system viewed at different scales.

Renormalization group techniques are applied to the action. The issue is that in a non-linear system, there are multiple time and space scales. Thus, it is necessary to use a standard singular perturbation analysis to arrive at a renormalized action that takes into account new critical points at longer time scales and longer length scales. The action changes, and a **renormalized action** is obtained as the result. This introduces a diffusion term since on longer time scales, reactions look like reaction-diffusion networks. This is the action required to obtain finite results from a non-linear system. As time goes to infinity, the propagator approximates the propagator of a Brownian motion, and the diffusion limit of a random walk. The key result is that at long timescales, the neural network dynamics looks like those of a reaction-diffusion system (which is important in determining network criticality events).

## 4. System Criticality
### 4a. Reggeon Field Theory

The renormalized action is the same as the action in Reggeon Field Theory. Reggeon Field Theory (1975)[86] is an extension of Regge theory, articulated by Regge (1959) and applied by Vladimir Gribov (1960s) in high-energy particle physics, in the Regge theory of strong interactions.[87] The behavior of a Reggeon Field Theory system corresponds to both branching and propagation (the crucial criticality in a neural system). Such systems have a universal non-equilibrium phase transition, in a class called directed percolation.

### 4b. Directed Percolation Phase Transition

Directed percolation is unidirectional percolation through a graph or network. A universal phase transition in non-linear neural networks is a directed percolated phase transition (DP phase transition).[88] DP phase transitions are non-equilibrium phase transitions, a universality class of directed percolation, which play a similar role in non-equilibrium systems as the Ising model in equilibrium statistical physics (i.e. serving as a classification model for system energy and interactions).[89] When balanced or equilibrium states of a neural network are destabilized, a DP phase transition can arise. Away from critical points (in which the system is stable and has low firing rates), a mean field theory such as Wilson-Cowan equations adequately describes the action and the system behavior. However, a different explanatory mechanism such as the DP phase transition is needed to describe what happens in system criticality.

The emergence condition for a DP phase transition is analogous to the Ginzburg condition in superconductivity (a boundary condition regarding the thickness of the superconducting plate).[90] The phase transition occurs at a threshold that is the combination of the level of excitation in the network and the diffusion spread (the length of the diffusion spread of the activity). An upper critical dimension is obtained, just as in superconductivity. Depending on the space parameters of the network, there could be critical branching, or branching plus aggregation, which is directed percolation. Branching is avalanches of spiking neurons, and the number of avalanches scales as a power law. Thus, power laws can be seen in the system dynamics. The effects of all the nodes in the network are captured in the DP phase transition formulation and provide the statistical activity (Brownian motion), which articulates system criticality.

## 5. Optimal Control Theory

Optimal Control Theory (OCT) is a set of mathematical optimization methods that can be used for the purpose of managing physical systems with quantitative structures, in the discipline of control engineering.[91] The statistical neural field theory derived here can be extended into an optimal control model. In particular, the action of an optimal control theory can be expressed as a path integral on the dynamical system.[92] A control theory can be externally applied as an action on the system, just as any intrinsically-arising action such as a signaling cascade is an action on the system. The orchestration mechanism is the path integral of the action.

In the brain's neural networks, a measure such as synaptic plasticity might be added to obtain a modified action as a feedback and control mechanism.[93] This is a way that some of Per Bak's self-criticality ideas could be embodied in management of critical moments in the network. The goal could be to produce a self-tuning system that is orchestrated by the synaptic plasticity operator. The system could self-adjust so it reaches (or avoids) criticality on shorter timescales with synaptic plasticity as an operator that either facilitates or depresses the action. With an operator as opposed to a constant function, the Hamiltonian would be more complex, so the underlying action would change, but not the renormalized action. The DP phase transition structure could likely persist and be used as a metric for optimal system control.

**Appendix 2**

**Model Field Theory #2: The Spin Glass Model (Wolynes)**

**Figure A2**. Major Steps in the Derivation of the Spin Glass Model for Energy Landscapes.

|  | **Problem/Requirement** | **Tools/Techniques** | **Results** |
|---|---|---|---|
| A | Description of the system at equilibrium | • General characterization<br>• Random (Gaussian) guesses at total system energy | • Random Energy Model |
| B | Description of the system at criticality | • Effective Field Theory | • P-spherical Spin Glass Model |
| 1a | Overcome glassy state with a glass transition (phase transition) | • Energy function (Hamiltonian)<br>• Spin glass (metastable system) that can be solved analytically | • System converges on a solution (a folded protein or a trained deep learning system that can accurately classify new test data) |
| 1b | Levinthal Paradox (proteins cannot be folding randomly) | • Structural contacts between regions of the protein | • Rugged convex energy funnel converges on a solution, a folded protein |
| 2 | A tool to understand system criticality<br>A model to detect Phase Transition | • Energy-entropy trade-off (information-theoretic system energetics) | • Identify/predict conditions for system phase transition, for example at a critical temperature (Tc) system where energy and entropy converge |
| 3 | Efficient system design to optimize goal-directed work | • Minimize loss and cross-entropy<br>• Energy function (Hamiltonian) | • Efficient (i.e. minimal) information encoding system |

**Model Field Theory #2: The Spin Glass Model (Wolynes)**

**Methods**
This section examines two proposals for field theory application in the areas of protein folding and deep learning networks, the Random Energy Model (basic) and Spin Glasses (sophisticated). Regarding protein folding, the summary discusses work by Peter G. Wolynes, a researcher in the Chemistry Department at Rice University, who proposed a spin glass model as an explanation for how proteins fold. The main paper outlining the idea is "Funnels, pathways, and the energy landscape of protein folding: a synthesis" (1995).[94] Also related is "The energy landscapes and motions of proteins" (1991),[95] and a Nature article considering the Random Energy Model and spin glasses (1997).[96] Regarding deep learning networks, sources include NYU machine learning expert Yann LeCun's research,[97] Charles Martin's summary of recent developments,[98] and Max Tegmark and Harry Lin's work regarding the application of physics principles to deep learning.[99,100]

**A. Basic Model: The Random Energy Model and Real Glasses**
Just as the Wilson-Cowan equations provide a good general characterization of a biological neural network through the mean field theory approach, the Random Energy Model fulfills a similar function for the spin glass model. The Random Energy Model provides a basic description of the system, and the Spin Glass Model is a more complicated extension of the Random Energy Model which can be solved exactly. The Random Energy Model makes random (Gaussian) guesses of the probability distribution of the total energy in the system.

In statistical physics, the Random Energy Model is typically applied as a basic model for analyzing systems with quenched disorder (disorder is frozen (quenched) into the system). The Random Energy Model was introduced as a possible model for glasses since glasses are amorphous (not regularly-ordered) systems. Classically, a glass is made by applying fire to sand. Most liquids freeze when they are cooled (e.g. water). However, when a liquid is supercooled it becomes a glass. This means that it gets viscous and only becomes amorphously solid (even if appearing fully solid in everyday use). All liquids can be made into glasses, if they are supercooled fast enough.

### The Glass Transition

The transition to glass is not a normal and complete phase transition (as in the case of particles to a gaseous state) because the atoms are held suspended, or quenched. In the transition to glass, the arrangement of atoms is amorphous (i.e. ill-defined), but not completely random. Different cooling rates and annealing techniques produce different glassy states (glass becoming either brittle or stiff). Energetics (the relation of energy and entropy within a system) and especially entropy collapse may be implicated in bringing about the glass transition.[101] Entropy collapse in the form of a "Kauzmann entropy crisis" is a situation occurring at lower temperatures within systems. The system runs out of entropy, meaning that there is a high availability of disordered energy to do work, but it cannot be usefully applied, and a static phase transition results.[102]

### Limitations of the Random Energy Model

To obtain a more precise measure than the basic approximation of the Random Energy Model, a Hamiltonian can be used to sum the weighted energy probability of each node's state. The total energy in the system can be calculated, along with local minima and maxima, which can be used to indicate the system's critical points. The Random Energy Model does not have an organizing principle such as a Hamiltonian to identify when and where the system might converge. In a practical example, the Levinthal Paradox arises. The paradox is that if a protein were folding randomly (as explained by the Random Energy Model), it would take longer than the known lifetime of the universe, and never fold because the system would never reach convergence. In actuality, proteins fold in a matter of nanoseconds.

### Overcoming the Glass Transition Problem

The glass transition problem is explaining how a system that seems to be stuck in a flat and persistent glassy state can enter one in which the energy landscape funnels down to converge on a critical point in order to do useful work, folding a protein, for example. The first class of solutions proposed for addressing the glass transition remains within the physical chemistry domain in which the problem occurs, and suggests that there could be some sort of natural pattern recognition that occurs at very low temperatures. While this could account for some observed supercooling phenomena, it does not explain protein folding. A second class of solutions is posited from the information-theoretic domain of energy and entropy, especially to model the system as a spin glass. The notion is that the physical glass transition might be avoided entirely via energetics (with an information-theoretic entropy model of the system). The premise is that there is a trade-off between energy and entropy. When systems are designed with good energetics (appropriate information-theoretic use of energy and entropy), the flat glassy state can be overcome, with instead, the system ending up in a funneled convex energy landscape in which solutions converge. The Levinthal Paradox of flat glassy surfaces and vanishing gradients can be resolved with a funneled energy landscape with rugged convexity, which can be understood more technically as a trade-off between system energy and entropy.

**B. Advanced Model: P-spherical Spin Glass**
A *spin glass* is a disordered magnet that is a metastable system in which roughly half of its molecular bonds are spin-up, and half spin-down.[103] Spin refers to the magnetic moment of an atomic nucleus arising from the spin of the protons and neutrons. The term "glass" comes from an analogy between the magnetic disorder of atomic spins in the spin glass concept, and the positional disorder of atomic bonds in a conventional glass. A conventional glass is an amorphous solid in which the atomic bond structure is highly irregular, as compared with a crystal, for example, which has a uniform pattern of atomic bonds. The system is called a *p-spherical spin glass* because in computing the system, the spins, which usually have spin-up or spin-down values in a quantum system, are relaxed into real numbers, and a spherical constraint is applied so that they sum to one.[104] These values are the spherical coordinates of a point P, hence the name p-spherical sin glass. The reason to use a spin glass model is because the model is known and analytically solvable.

From an energy-entropy trade-off perspective, the idea is that there is a system that is being described with an energy function (a Hamiltonian), with some up-spins and some down-spins. The system will be in the lowest energy state when as many as possible up and down spins are paired (e.g. neutralized). A spin glass model is metastable at best in that there is no one final solution. The system is forever "frustrated" with constraints that cannot be satisfied due to the structure of its molecular bonds. The properties of a spin glass are analogous to those of a semiconductor; the metastability makes them perfect objects for manipulation in computational (electromagnetic) systems. Likewise, as mentioned in the Cowan section (Appendix 1), Hopfield's important innovation for deep learning networks was noticing that spin glasses (as an Ising model of ferromagnetism) are a physical system in which energy can be minimized as a computational feature.

**Wolynes Spin Glass Model**
The spin glass is a disordered ferromagnet, or orientational glass, in which the energy landscape can be directed and funneled. A spin glass can be created by giving the Random Energy Model a ground state. The system energy is random, except for one state, the ground state, which is an attractor state. Specifying the energy of the ground state causes the flat glassy surface to disappear and descend into an energy funnel, similar to that of a hurricane.[105]

***The Energy Landscape Theory of Protein Folding***
In the context of protein folding, the frustration in the system is minimized by providing connections, or structural contacts, between regions in the protein.[106] Attempting to induce a polymer to fold without any connections results in a flat energy landscape (and it would never fold, per the Levinthal Paradox). However, adding connections to the polymers allows the system to converge. As more connections are added to the protein, the energy landscape becomes increasingly funneled and rugged, and converges. The system becomes convex, and produces an energy landscape that descends; a rugged and convex spin glass energy landscape. Using a spin glass version of the Random Energy Model also allows the foldability of a protein to be computed (measured information-theoretically).[107]

***Deep Learning applications of the Spin Glass Model: Loss Optimization***

Many systems have problems that can be structured in the form of an energy landscape that converges (i.e. an energy funnel), and these can be evaluated with a spin glass model. Continuing in the biological domain, in addition to proteins, spin glass models have been applied to other biomolecule conformations and interactions,[108] and crystal nucleation.[109] In the brain, the spin glass model might explain the collective behavior of neural networks. Specifically, the statistical mechanics of infinite-range Ising spin-glass Hamiltonians provides one explanatory model for collective neural behavior (that is methodologically similar to Cowan's Statistical Neural Field Theory).[110] Relating spin glass models to neural networks in the brain suggests their further applicability to neural networks in computation such as deep learning systems.

Spin glass models are typically applied to deep learning systems for optimization, in order to create efficient *loss functions* that reduce combinatorial complexity. The network can run more expediently if it can distinguish relevant features quickly instead of trying all possible permutations. Hinton's backpropagation of errors method (1986) was a key advance in loss optimization methods for deep learning networks and continues to be a focal point for research in the field.[111] Formulating loss optimization as an energy functions with a Hamiltonian is a standard technique in deep learning systems.

One method that applies spin glass models to deep learning engages research regarding the complexity of spherical spin-glass models.[112] This research formulates an identity between the critical values of random Hamiltonians, and the eigenvalues of random matrix ensembles (Gaussian ensembles). The identity is then used to calculate the ground state energy of the system, and the subsequent levels of energy in the system. These values can be used to determine the bottom or minimal energy landscape, and this structure is consistent with the transition from a glass to spin glass system. The formulation can be applied to loss optimization in deep learning networks. In an empirical case, the lowest critical values of the Hamiltonians formed a layered structure and were located in a well-defined band that was lower-bounded by the global minimum.[113] The result was that the loss function of the neural network displayed a similar landscape to that of the Hamiltonian in a spin-glass model (e.g. funneled and directable).

The spin glass model has also been used for loss optimization in deep learning in other prominent recent advances such as dark knowledge and adversarial networks. *Dark knowledge* is a compositing technique in which the predictions of several model runs are compressed into a single model to generate more efficient results.[114] In spin glass information-theoretic terms, the dark knowledge model continues to have the same entropy as network layers are added and compressed, but the loss function keeps improving, decreasing the error rate and delivering better results. The same structural point is true with adversarial networks. *Adversarial networks* is a self-play method in which there two networks. An adversary network generates false data and a discriminator network tries to learn whether the data are false, not by changing the structure of the neural network, but again by working with the convergence efficiency of the loss function.[115]

**How are energy and entropy related in a system?**

Figure A2a is one representation of how the energy and entropy are related in a system.[116] An image of the canonical phase transition between a solid, liquid, and gas phase based on temperature and pressure appears on the left side for comparison (with the critical Temperature Tc). In the Energy-Entropy trade-off image on the right, phase transition also occurs at the critical Temperature at which the energy and the entropy converge, when the entropy has collapsed to be the same as (converge with) the free energy in the system. The free energy and the entropy are the same. An intuitive interpretation of this formulation could be that when system entropy (order decay) and energy are equal, a phase transition occurs. There are two temperature phases, one above and one below the critical temperature (Tc) at which the phase transition occurs. Below Tc, the critical temperature, the entropy vanishes and the system thermodynamics are dominated by a small set of configurations. Above Tc, there is an exponentially large number of possible system configurations, and at this dispersed energy, the Boltzmann measure of entropy is roughly equally distributed between the configurations. In such a system, the "annealed" entropy density is a function of the energy density. A phase transition is defined as a non-analyticity of the free energy density. The system is metastable at the phase transition moment because it is balanced between energy and entropy (between a low and high number of configurations). The phase transition occurs at a certain convergence of parameters.

**Figure A2a**. Phase Transitions.

| Standard Temperature-Pressure Phase Transition Diagram for a Fluid. | Energy-Entropy Trade-off. (Mézard Fig. 5.3, Pp. 98-99) |

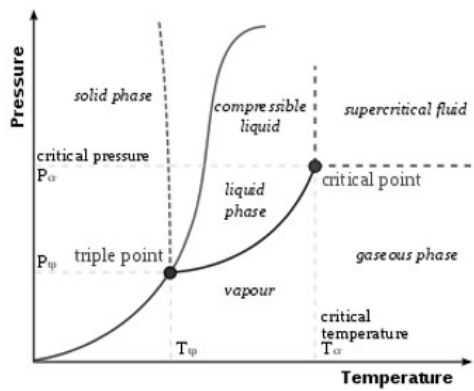 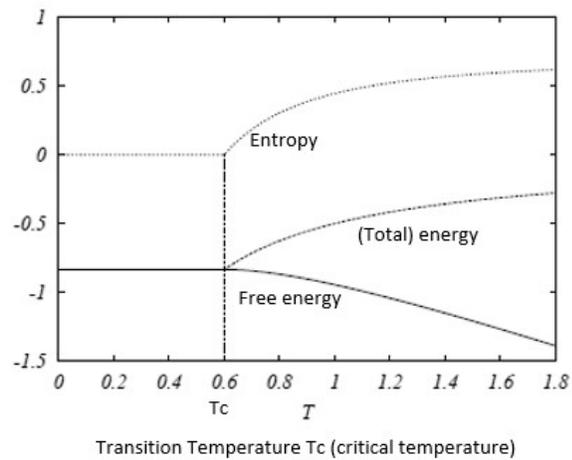

Energy-related terms such as the Hamiltonian are important for their role in describing the trade-off between energy and entropy. Systems tend to move toward a higher entropy state over time, whether the universe, a child's bedroom, or a smart network system. The laws of thermodynamics hold in the sense that it would not be possible to transfer free energy without a corresponding energy expenditure in the form of the entropy that the system must spend. The trade-off between energy and entropy in smart network domains is a relationship between the terms that is maintained as the system operates, including at criticality and phase transition.

Energetic principles may also be considered in the sense that entropy measures how the energy of a system is dispersed. For example, when water is ice, the system has a high degree of order, but when there is a phase transition and the water is water, entropy goes up because there is less order in the system (molecules float randomly and are not stored as ice crystals). The reason to think of smart network systems in terms of entropy and energy is that it casts them in information theoretic terms, which could be useful for future formulations. Energy and entropy are related as articulated in the maximum entropy principle: for a closed system with fixed energy and entropy, at equilibrium, energy is minimized and entropy is maximized (such as a pitcher of water versus an ice block).

**Deep Learning Network goal: Loss Optimization and Cross-Entropy Minimization**
The statistical mechanical approach to deep learning frames networks as a problem of optimizing an energy function. Loss optimization is the primary focus, but minimizing cross entropy is another important information-theoretic technique. Another way to use the system's energy function is with the goal of minimizing cross entropy. *Cross entropy* is a measure of the efficiency of an information encoding system. For example, in an autonomous driving system that is only distinguishing between vehicles and non-vehicles, a binary yes-no suffices for information encoding. On the other hand, if a smart city traffic sensor system is identifying vehicles by make and model, more bits will be required to efficiently encode the information required to label the data. In the first example, cross entropy is minimized if the simplest binary encoding scheme is used. Cross entropy is a mechanism that can be used to quantify and compare the difference between two probability distributions and their efficiency in information encoding.

The analogy in statistical physics is that free energy is being minimized.[117] A deep learning network is structured as an energy function (Hamiltonian) and the goal is to minimize the cross entropy.[118] The energy function is used in the deep learning network as an overall organizing principle to describe the activity of the system processing units (perceptrons). Specifically, the energy function (Hamiltonian) is defined as a combination of the parameterized summing of Gaussian random variables (processing nodes) and their weights.

# Appendix 3

**Figure A3**. Phases in the Progression to Digital Reality.

| | Domain | Digitization (existing in a digital format) | Modularization (standardized units, distributable globally) | Marketization (credit and remuneration matched with contribution) | Computability (instantiated in an automatically calculated format) |
|---|---|---|---|---|---|
| 1 | Information | Yes | Yes, private silos: scientific, corporate, governmental, consumer | Data Markets | Undefined |
| 2 | Entertainment | Yes | Yes | Yes | Yes |
| 3 | News | Yes | Yes | Yes | Yes |
| 4 | Knowledge | Yes, partially | Wikipedia, Coursera, Khan Academy | Knowledge Markets | Undefined |
| 5 | Computation | Yes | Smart contracts | Computation Markets (Ethereum) | Undefined |
| 6 | Publications | Digital format | Research Models have own DOIs | Research Citations | Computable (Wolfram) (No) |
| 7 | Science | Technophysics as a data science method | Concepts, methods, approaches, formulae | Science Markets, Citizen Science | Auto-hypothesis generation |
| 8 | Health | EMRs, genomes, preventive medicine | Global Health Care Equivalency Markets | | Big health data, disease causality |
| 9 | Assets | Blockchain-based digital asset registries | Yes, partially | Real-time Balance Sheets, Programmable Risk | Undefined |
| 10 | Intelligence | Social media, life-logging, mind-file backup, uploading | Brain as a DAC | Idea Markets | Cloudminds |

*Legend: colored-in boxes = achieved; otherwise undefined: no clear vision of what this is yet*

**The world's science methods are modularized and globally distributable**
Technophysics includes the notion of *science modules* (standardized usable units), in the same way that Coursera is *knowledge modules*, standardized units of information required for the understanding of a certain topic packaged into a consumable unit and globally distributable. Smart network field theory could help in the specification of science modules that include the concept, approach, method, and formalism, possibly applied in multiple fields, packaged for deployment as a technophysics science module. The idea is to have a standardized toolkit that is readily mobilizable by persons without deep knowledge in the field to new contexts. Anyone worldwide can use a Schrödinger wave equation to test in other problem domains. The premise is that the conceptual intuition and the formal application of the scientific method are readily-deployable by persons outside the field, for example in smart network field theory modules.